\documentclass[12pt,preprint]{aastex}
\usepackage{epsfig}
\usepackage{changebar}

\bibpunct{(}{)}{;}{a}{}{,}

\slugcomment{2003ApJS..149..265R}

\shorttitle{The {\sc Acbar} Instrument}
\shortauthors{Runyan et al.}

\usepackage{amsmath}
\usepackage{amssymb}

\begin{document}

\title{ACBAR: The Arcminute Cosmology Bolometer Array Receiver}



\author{M.C. Runyan\altaffilmark{1,2},
P.A.R. Ade\altaffilmark{3},
R.S. Bhatia\altaffilmark{4},
J.J. Bock\altaffilmark{5}, 
M.D. Daub \altaffilmark{6}, 
J.H. Goldstein\altaffilmark{7,8}, 
C.V. Haynes\altaffilmark{3},
W.L. Holzapfel \altaffilmark{6},
C.L. Kuo\altaffilmark{6,9}, 
A.E. Lange\altaffilmark{1},
J. Leong\altaffilmark{7},
M. Lueker\altaffilmark{6},
M. Newcomb\altaffilmark{6},
J.B. Peterson\altaffilmark{10}, 
C. Reichardt\altaffilmark{1},
J. Ruhl\altaffilmark{7}, 
G. Sirbi\altaffilmark{4},
E. Torbet\altaffilmark{8},
C. Tucker\altaffilmark{3},
A.D. Turner\altaffilmark{5},
D. Woolsey\altaffilmark{6}}

\altaffiltext{1}{Department of Physics, Math, and Astronomy, California Institute 
of Technology, Pasadena, CA 91125}
\altaffiltext{2}{Current address: Enrico Fermi Institute, University of Chicago, LASR-132,
Chicago, IL 60637: mcrunyan@grizelda.uchicago.edu}
\altaffiltext{3}{Department of Physics and Astronomy, Cardiff University, CF24 3YB Wales, UK}
\altaffiltext{4}{European Space Agency, 2200 AG Noordwijk, The Netherlands}
\altaffiltext{5}{Jet Propulsion Laboratory, Pasadena, CA 91125}
\altaffiltext{6}{Department of Physics, University of California at Berkeley, Berkeley, CA 94720}
\altaffiltext{7}{Department of Physics, Case Western Reserve University, Cleveland, OH 44106}
\altaffiltext{8}{Department of Physics, University of California, Santa Barbara, CA 93106}
\altaffiltext{9}{Department of Astronomy, University of California at Berkeley, Berkeley, CA 94720}
\altaffiltext{10}{Department of Physics, Carnegie Mellon University, Pittsburgh, PA 15213}

\begin{abstract}
We describe the Arcminute Cosmology Bolometer Array Receiver ({\sc Acbar}); a multifrequency
millimeter-wave receiver designed for observations of the Cosmic Microwave Background 
(CMB) and the Sunyaev-Zel'dovich effect in clusters of galaxies.  
The {\sc Acbar} focal plane consists of a 16-pixel, background-limited, 240 mK 
bolometer array that can be configured to observe simultaneously  
at 150, 220, 280, and 350 GHz.  With $4-5^\prime$ FWHM beams and 
a $3^\circ$ azimuth chop, {\sc Acbar} is sensitive to a wide range of angular scales. 
{\sc Acbar} was installed
on the 2 m Viper telescope at the South Pole in January 2001. 
We describe the design of the instrument and its performance during the 2001 
and 2002 observing seasons.
\end{abstract}


\section{Introduction}\label{sec:intro}

The Arcminute Cosmology Bolometer Array Receiver ({\sc Acbar}) 
was designed to make sensitive, high-resolution measurements of the microwave sky 
at multiple millimeter wavelengths.  
Observations of the cosmic microwave background (CMB) can provide a wealth of information
about the Universe. Measurement of the angular power spectrum of the CMB 
allows us to determine the values of the
cosmological parameters (such as the age, density, and composition of the Universe)
within the context of models of the early universe.

The angular power spectrum of the CMB 
is exponentially damped on small angular scales
by photon diffusion \citep{silk68,white01}.
On very small scales ($\lesssim 5^\prime$) it is believed the CMB power spectrum
will be dominated by sources of secondary anisotropy. One such source is
the scattering of
CMB photons by the hot plasma in foreground clusters of galaxies 
[see, for example, \citet{holder99b} or \citet{cooray02}]
known as the Sunyaev-Zel'dovich (SZ) effect \citep{sz,birkinshaw99}.  
Observations with the CBI and BIMA may have already detected this SZ
power spectrum \citep{bond02,dawson02,komatsu02,mason03}.  
Multifrequency observations should allow
the separation of the primary and SZ power spectra through the unique frequency 
dependence of the SZ effect.

The field of millimeter-wave bolometer observations of the CMB has a long
pedigree with both ground-based [IRTF \citep{meyer83}, QMC-Oregon \citep{ade84},
IRAM \citep{kreysa87}, SEST/OASI \citep{pizzo95}, SuZIE \citep{holzapfel97a,
mauskopf1835}, Python \citep{platt97}, SCUBA \citep{holland99}, and Diabolo \citep{benoit00}]
and balloon-borne experiments [TopHat \citep{cheng94}, PRONAOS \citep{lamarre94}, 
MSAM \citep{fixsenmsam}, MAXIMA \citep{lee99},
and BOOMERANG \citep{crill02}].  The {\sc Acbar} experiment drew upon
much of this well-developed technology, thus enabling rapid construction
and deployment of a very sensitive instrument.

The {\sc Acbar} focal plane is a background limited, 16-pixel,
240 mK bolometer array that is configurable to observe simultaneously 
at 150, 220, 280, and 350 GHz with $4-5^\prime$ resolution.  Observations
at these frequencies span the peak intensities of CMB anisotropies
and the SZ effects while avoiding foreground contaminants, such as
dust and radio point sources \citep{tegmark96}.
The technology, frequency coverage, and angular resolution
of {\sc Acbar} are very similar to those being used in the {\it Planck} HFI.  
{\sc Acbar} was installed on the 2 m Viper telescope at the South Pole in January 2001.  
{\sc Acbar} exploits the excellent polar atmospheric
conditions and large $3^\circ$ telescope chop to measure angular scales from
$\ell\sim 75$ to $\sim3000$. 

This paper describes the details of the instrumental design and performance of
{\sc Acbar}.  The telescope and instrument were upgraded in December 2001 and we 
describe the performance during both the 2001 and 2002 observing seasons.
We present the CMB power spectrum measured by {\sc Acbar} in \citet{kuo02}
and constraints on cosmological parameters derived from this data
in \citet{goldstein02}.  We describe pointed observations
of nearby clusters in \citet{gomez02} as well as a survey for clusters
of galaxies with the SZ effect in \citet{runyan03b}.

We describe the physical 
environment at the South Pole in \S\ref{sec:enviro} and discuss the
details of the telescope and focal plane optics in \S\ref{sec:optics}.  
The cryogenics employed to cool the detectors to 240 mK are described
in \S\ref{sec:cryo} and the
low-noise electronics are discussed 
in \S\ref{sec:electronics}.  In \S\ref{sec:obs} and
\S\ref{sec:calib}, we describe the observations during 2001 and 2002 as well as
the calibration of the instrument.  
We present the achieved sensitivity of the
instrument in \S\ref{sec:noise} and our conclusions in \S\ref{sec:conc}.

\section{The South Pole Environment}\label{sec:enviro}

{\sc Acbar} observes from the Viper telescope located at the Southern geographic pole in
Antarctica.  The South Pole provides a unique platform for terrestrial far-infrared
observations.  The Amundsen-Scott Station is located atop the
Ross ice shelf at an elevation of $9,300^\prime$ which reduces the column depth of
atmosphere above the telescope.  The pressure elevation at the Pole can exceed
$11,000^\prime$ due to the 
thinning of the polar atmosphere from the extreme cold and
the bulging of the atmosphere at the equator.  The
cold temperature freezes out most of the precipitable 
water vapor from the atmosphere, greatly
reducing emission and absorption at millimeter wavelengths.  
The ambient temperature averages near $-80^\circ$F
in the austral winter with a precipitable water vapor less than 0.32 mm $\sim75\%$
of the time
\citep{lane}.  In addition, the atmosphere is stable for long periods of time 
\citep{peterson02}
punctuated by short periods of poor weather (usually associated with a warming 
of the ambient temperature) and has no diurnal variation.  
The entire Southern celestial hemisphere is available
year round allowing long, continuous observations of a field.  
Combined with
a well established research infrastructure, these attributes
make the South Pole arguably one of the best locations on the planet for
millimeter-wave observations.

\section{Optics}\label{sec:optics}

\subsection{Telescope Optics}\label{ssec:teloptics}

The Viper telescope is located less than one kilometer from the South Geographic Pole, in close
proximity to the Amundsen-Scott South Pole station.
Viper is an off-axis aplanatic Gregorian
telescope with a re-imaging tertiary mirror to reduce the effective focal length.  
Viper has a 2 m diameter primary mirror and additional 0.5 m 
radius reflective skirt to
reflect primary spillover to the sky.  
There is a chopping flat mirror located at the image of the
primary formed by the secondary mirror which sweeps the beams approximately $3^\circ$ on the
sky with little modulation of the beams on the primary (of order a few cm).

\begin{figure}
\centering
\plotone{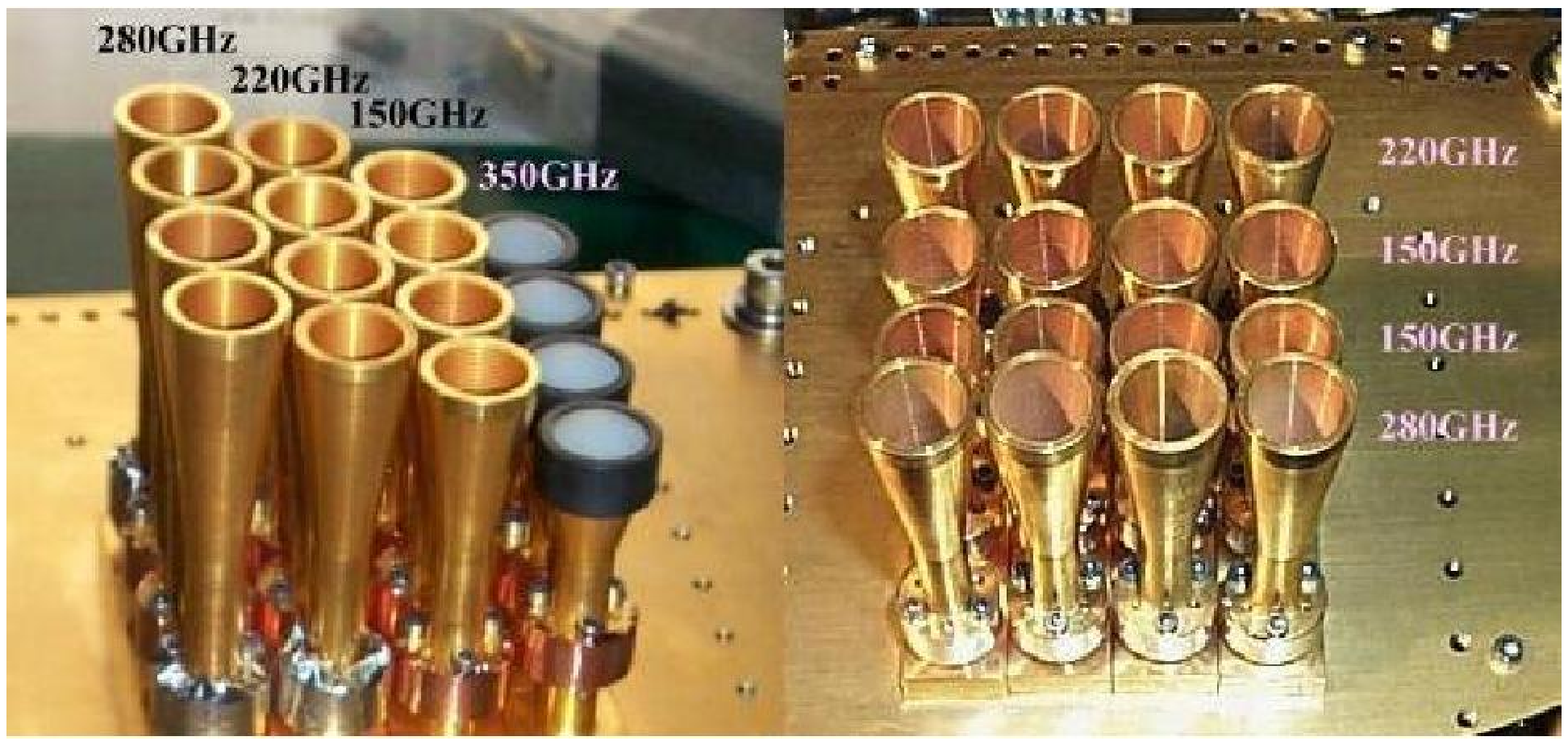}
\caption[Photos of {\sc Acbar} Feed Arrays for 2001 and 2002]{Images of the {\sc Acbar} 
focal plane layout for 2001 (left) and 2002 (right).} 
\label{fig:layouts}
\end{figure}

\begin{figure}
\rotatebox{90}{\plotone{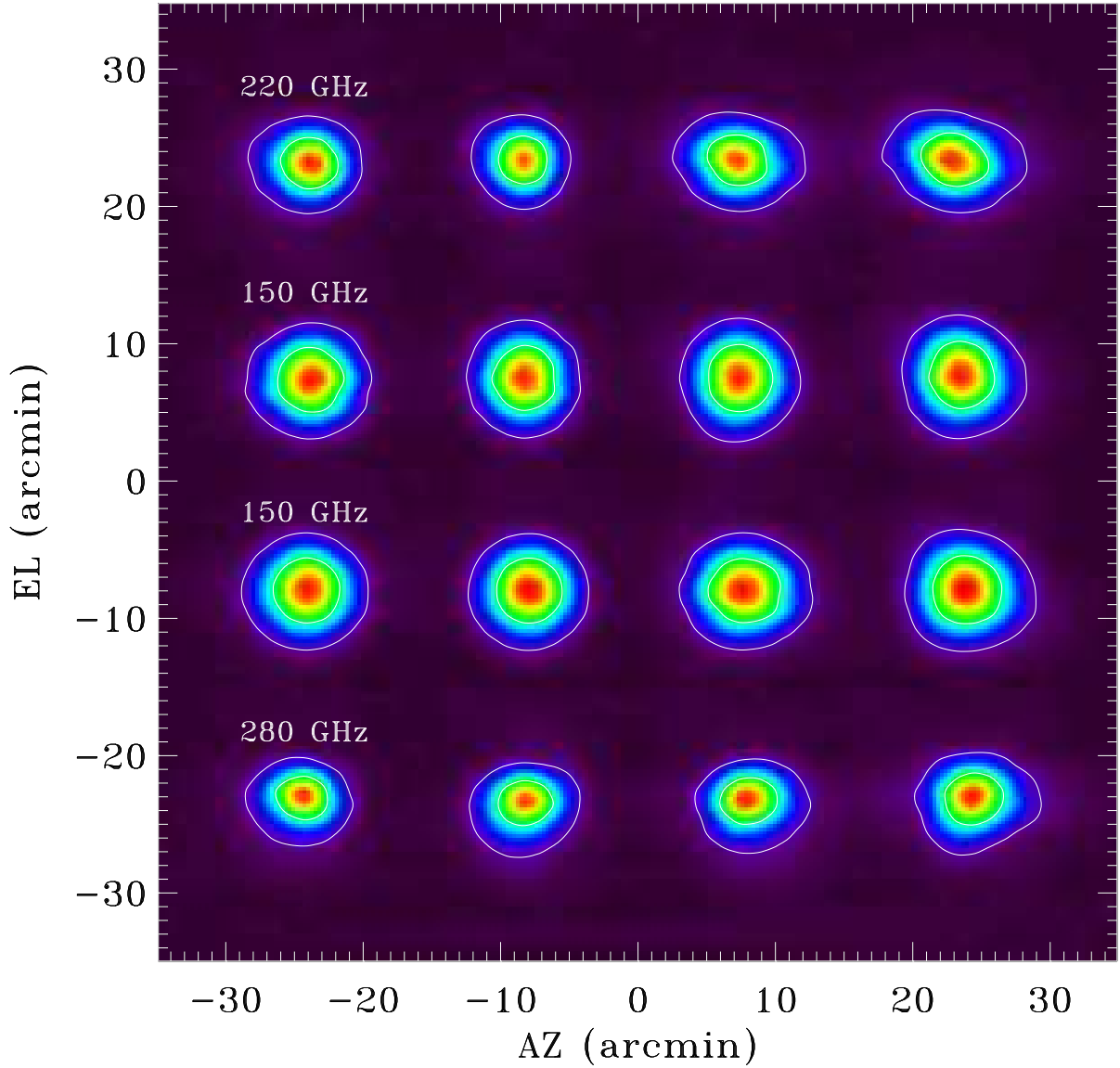}}
\caption[Raster Map of Venus in 2002]{Raster map of Venus at an elevation of 
$19.8^\circ$ taken on 2002 September 22.  The angular diameter of Venus was 
$37^{\prime\prime}$ on this date and the map has been smoothed with a $1^\prime$
FWHM Gaussian.  The white contours correspond to power levels
of 10\% and 50\%.}
\label{fig:venus}
\end{figure}

The re-imaging tertiary on the Viper telescope was originally designed for low frequency (40
GHz) observations of the CMB with a 2-pixel array \citep{peterson00}.
The tertiary was
redesigned using geometrical ray-tracing software
to provide sufficient optical quality across a large field of view ($\sim1~{\rm deg}^2$)
at {\sc Acbar}'s higher frequencies. With the new tertiary, 
the telescope has an effective focal length of
3.44 m, resulting in a plate scale of $\sim 1^\prime/{\rm mm}$.  
The 16 mm
separation of feeds on the {\sc Acbar} focal plane (see Figure \ref{fig:layouts})
produces a $4\times 4$ array of beams 
separated by $\sim16^\prime$ on the sky (see Figure \ref{fig:venus}).

\begin{figure}
\centering
\plotone{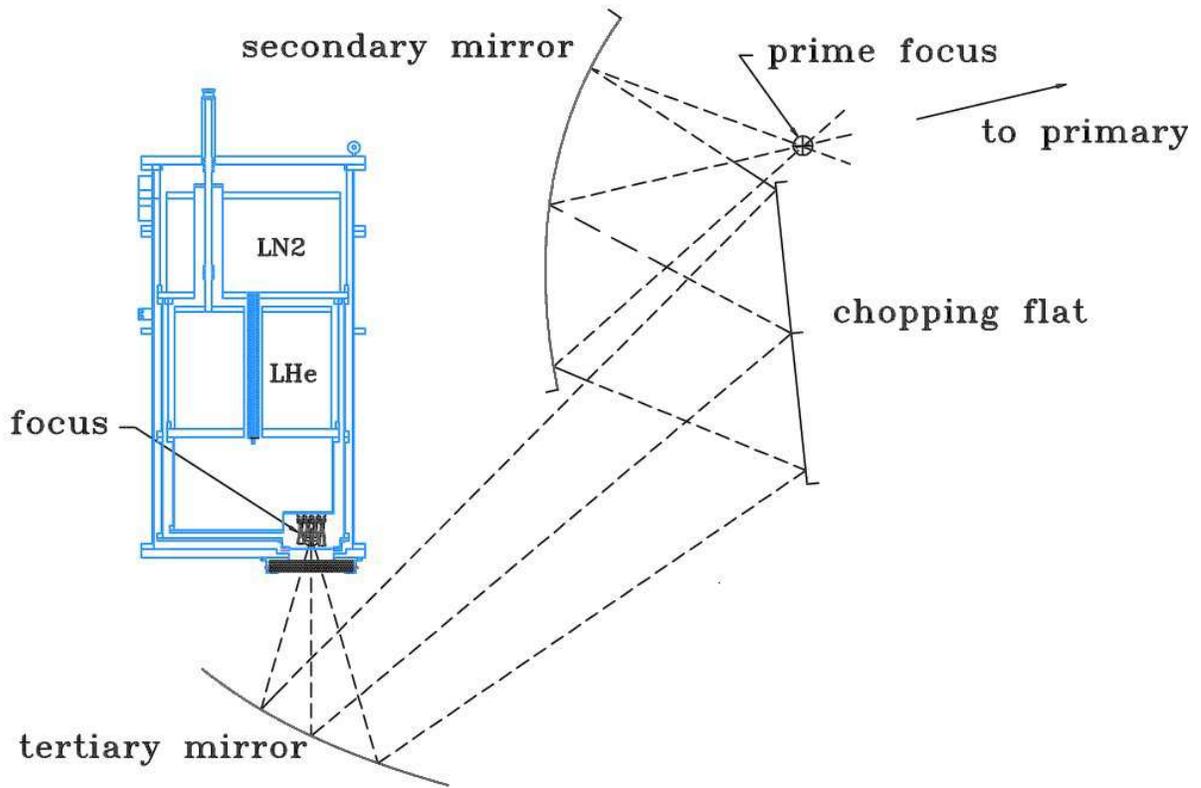}
\caption[Telescope Optics Arrangement]{Most of the Viper telescope optics 
along with the {\sc Acbar} cryostat.  The light rays are the geometrical optic rays for 
full illumination of the 2 m primary and are shown to illustrate the tight clearance 
around the prime focus.} 
\label{fig:viper}
\end{figure}

Figure \ref{fig:viper} shows a schematic of a subsection the Viper telescope optics along
with the {\sc Acbar} receiver.  The extreme geometric optical rays are drawn to
illustrate the tight clearances of the system; the actual Gaussian beam widths are much more
narrow than the lines drawn.  With full illumination of the 2 m primary, the focal plane is
fed at $f/1.7$ in the geometric optics limit.  The primary illumination
varies with frequency, however, and so the effective half-power $f/\#$ of 
the beams vary from $f/4.6$ at 150 GHz to $f/11.1$ at 350 GHz.
The dewar is mounted on a sliding
pillow-block mount with a linear focus actuator that controls the distance between the 
dewar and tertiary mirror.

The chopping flat mirror sweeps the beams across the sky approximately $1^\circ$ for every
$2.2^\circ$ of chopper rotation.  The position of the chopper is read out with an 
angular encoder and is controlled to trace a triangular chop with a PID loop.  
The triangle wave results in a constant-velocity chop across the sky.
The chopper encoder waveform is fed into the data acquisition system as an
additional signal and sampled at the same rate as the bolometers ($\sim2400$ Hz).
When observing the CMB
we sweep the beams approximately $3^\circ$ across
the sky in about a second, which effectively freezes the atmosphere.  The chopper
speed is chosen so that the beams will take multiple detector time constants to cross a
point source.  We employed a chopper speed of 0.7 Hz in 2001 and 0.3 Hz in 2002.  When 
observing known clusters we shorten the chop length to $1.5^\circ$ to concentrate the
observing time on the cluster as well as increase the chopper frequency to keep the
scan velocity the same as for CMB observations.

During the 2001 season, we used an RVDT angular 
encoder\footnote{Schaevitz Sensor, model \#R30A} as
both the control signal and recorded position signal.  
The RVDT suffered from two features that
were the dominant sources of pointing error in 2001. 
First, a drift in the zero-point voltage of
the encoder caused the position on the sky corresponding to a fixed
encoder voltage to change with time.  
This drift was slow (about 10$^\prime$ per month) and we were able to fit and
correct for it
with frequent observations of galactic sources.  The second more serious
problem was the fluctuation of encoder gain of about 8\%.  
The gain of the encoder was roughly bimodal, falling into a
high-gain or low-gain state.  To determine the gain state of the encoder we compare the
separation of a bright object (in volts) between two adjacent channels.  
To remedy these problems, we installed
an absolute optical encoder\footnote{Gurley Precision Instruments, model \#A25S} 
for 2002.
The RMS pointing error during 2001 was approximately 1.3$^\prime$ and the RMS for 2002 was
30$^{\prime\prime}$.

The telescope is enclosed in a large, 
conical ground shield that reflects telescope spillover
to the sky.  The ground shield also 
blocks emission from elevations below $\sim25^\circ$.
This reduces optical loading and the effects of modulated sidelobes.  
One section of the ground shield lowers to allow observations of low-elevation
sources; this is necessary to observe planets which do not rise higher than $\sim30^\circ$
above the horizon at the pole.  Fortunately, most of the region of the Southern
Hemisphere with the lowest dust contrast is located at $EL>25^\circ$ at the pole.

A servo-controlled
PID loop controls the positioning of the telescope azimuth and elevation
with power supplied by linear servo
amplifiers.  All temperature-sensitive components of the telescope are heated
to $\sim300$ K.  The mirrors are equipped with heaters for sublimating the
thin layer of ice that accumulates over time.  Blowing snow collects on most of the
mirrors and must be cleaned off daily; this prevents snow from 
contributing to chopper synchronous offsets as well as
attenuating astrophysical signals, as discussed below in \S\ref{sec:offsets}.

We developed a telescope pointing model using frequent observations of both galactic and
extragalactic sources.  This allows us to reconstruct the position of each beam on 
the sky using the reported telescope 
AZ, EL, and chopper encoder positions.  The pointing model incorporates the
distance of the telescope from the geographic pole, the tilt of the azimuth ring, flexure of
the telescope with elevation, and the collimation offsets between the radio beams and
nominal telescope boresight position.  The measured AZ and EL chopper functions
are then used to translate the measured chopper position into an instantaneous beam position
for all 16 optical channels.  
Because of the proximity to the geographic pole, the telescope requires very little change
in elevation while tracking a source across the sky.

\subsection{Focal Plane Optics}\label{ssec:fpoptics}

The focal plane optics are designed to couple the {\sc Acbar} receiver to the Viper
telescope and produce diffraction-limited beams at 150 GHz.  The angular resolution at
higher frequencies is intentionally degraded to produce nearly matched beam sizes at all
frequencies.  
Figure \ref{fig:layouts} shows the layout of the {\sc Acbar} focal plane as configured for
observations in 2001 and
2002.  In 2001, the focal plane was arranged with common frequencies aligned in columns so
that each row observed at 150, 220, 280, and 350 GHz as the chopper swept across the sky.  
For 2002, the 350 GHz feeds were replaced 
with an additional set of 150 GHz feeds (because of their poor noise performance). 
The focal plane was arranged
with rows of common frequency to concentrate the declination extent of the 150 GHz channels
in 2002.  
The design of the {\sc Acbar} feed structure is based upon the {\it Planck} satellite prototype
design of \citet{feed} (see Figure \ref{fig:feedstructure}).  
The main optical elements of the focal plane are the
beam defining scalar feeds, expanding and reconcentrating conical feeds, filters, and
bolometric detectors.  Each of these will be described in the following sections.

\subsection{Scalar Feeds}\label{ssec:feeds}

{\sc Acbar}'s corrugated feeds are designed to produce single-moded,
nearly Gaussian beams with very low sidelobes.  This reduces optical loading from
telescope spillover and decreases offset signals from modulation of the spillover by
the chopping flat of Viper.  Gaussian illumination also results in the smallest beam sizes
from an aperture.
Figure \ref{fig:feedstructure} shows the {\sc Acbar} 150 GHz scalar
feed structure.  The feeds were fabricated by Thomas Keating, Ltd.\footnote{Billingshurst,
England, http://www.terahertz.co.uk} and performed well within
specifications (see Table \ref{tbl:beamparams}).  The geometry of the conical section of the
scalar feeds (aperture diameter and aspect ratio) is designed to produce primary mirror beam
waists that scale with wavelength.  We modeled the
expected beam patterns by assuming balanced hybrid conditions at the aperture and used 
a Gauss-Laguerre expansion of the $HE_{11}$ mode as described in
\citet{wylde}.  The aperture and length dimensions of all four feeds are also
listed in Table \ref{tbl:beamparams}.  
The corrugation geometry within the conical section is designed in accordance
with \citet{clarricoats84}; 
the corrugation depth is $\lambda/4$ and the groove pitch is three
grooves per wavelength to preserve the desirable $HE_{11}$ mode.  We employ a meniscus
lens to shorten the 350 GHz feeds \citep{clarricoats69b}; this results in an increase
in sidelobe level but the shorter feeds are much less expensive to fabricate.  The
lens is anti-reflection corrugated with small holes to reduce surface reflections
\citep{kildal84}.

The feeds 
transition from corrugated to smooth-walled waveguide after the beam-defining
conical section.  
Because {\sc Acbar} is not sensitive to polarization, we are not concerned about instrumental
cross-polarization from mode conversion in the smooth-walled section. 
Smooth-walled structures are much
easier to fabricate and significantly less expensive than corrugated waveguide.  
The transition to smooth-walled
waveguide occurs in the throat section of the scalar feeds
and gradually converts the $\lambda/3$
pitch corrugations to smooth wall without abrupt changes in waveguide impedance.
We implement the throat prescription of \citet{zhang93}
which varies both the corrugation
thickness and depth through the throat.  

The beam patterns from the dewar were measured with a chopped thermal source during 
instrument integration at U.C. Berkeley in the Fall of 2000.  
Table \ref{tbl:beamparams} lists the measured Gaussian FWHM of the {\sc Acbar} beams exiting the
dewar as well as the model Gaussian widths.  The measured beam sizes from
the feeds agree well with the model-predicted Gaussian widths. 
The beam widths from the cryostat scale with wavelength and thus longer
wavelengths illuminate proportionally more of the primary.
In the absence of diffractive effects, scattering, and aberrations, 
this will produce matched beam sizes on the sky.

\subsection{Smooth-wall Section}\label{ssec:feedstruct}

Light enters the feed
structure through the beam defining scalar feed and encounters circular waveguide that 
high-pass filters the incoming light.  The length of the circular waveguide cutoff 
section is approximately three times the cutoff wavelength in order to fully 
attenuate low frequencies.  The sharp cutoffs of the waveguide edges
can be seen in the transmission spectra in Figure \ref{fig:szbands}.

After waveguide filtering, the smooth walled section re-expands to a diameter of several 
wavelengths before reaching the edge-defining and blocking filters.  This re-expansion is
necessary to make the guide wavelength equal to the free space wavelength; the 
metal-mesh filters are designed to operate in free space.  The metal-mesh filters are
separated from one another with polyurethane spacers that damp high-frequency leakage 
around the outside of the filters and standing modes between the filters.
There is a thermal break to
isolate the 4 K front-end of the feeds from the ultra-cold 240 mK 
detector side of the feed structure.

In general, the coupling between two conical feeds is poor because the phase
caps for the feeds have small radii with opposite curvature, resulting in a large phase
mismatch.  HDPE lenses at the apertures of both
conical feeds place a common beam waist midway between the two feed apertures and 
dramatically improve the coupling \citep{churchlens}.  
We tested the efficacy of the coupling lenses at 150 GHz by measuring the optical efficiency
of the feed structure with and without the lenses and found that the lenses improved the
relative optical efficiency by $\sim 40\%$.

The light is reconcentrated and fed into the detector cavity where it is absorbed by a
bolometer.  
The cavity entrance aperture and reflective backshort are both spaced
an odd number times $\lambda_0/4$ from the detector to produce 
constructive interference at the absorbing mesh. 
The optimal size of the entrance to the bolometer cavity was determined by gradually increasing
the diameter of the aperture and measuring the optical efficiency for each diameter.  
Expanding the aperture from the cutoff waveguide diameter to the full absorber diameter
resulted in a relative optical efficiency improvement of $\sim40\%$ at 150 GHz
and $\sim80\%$ at 350 GHz.
We believe the improvement in coupling with larger detector cavity aperture is caused by 
two effects: the guide wavelength is closer to free space and the beam suffers less 
diffraction at the cavity aperture.  

\subsection{Filters and Bands}\label{ssec:filters}

\begin{figure}
\centering
\plotone{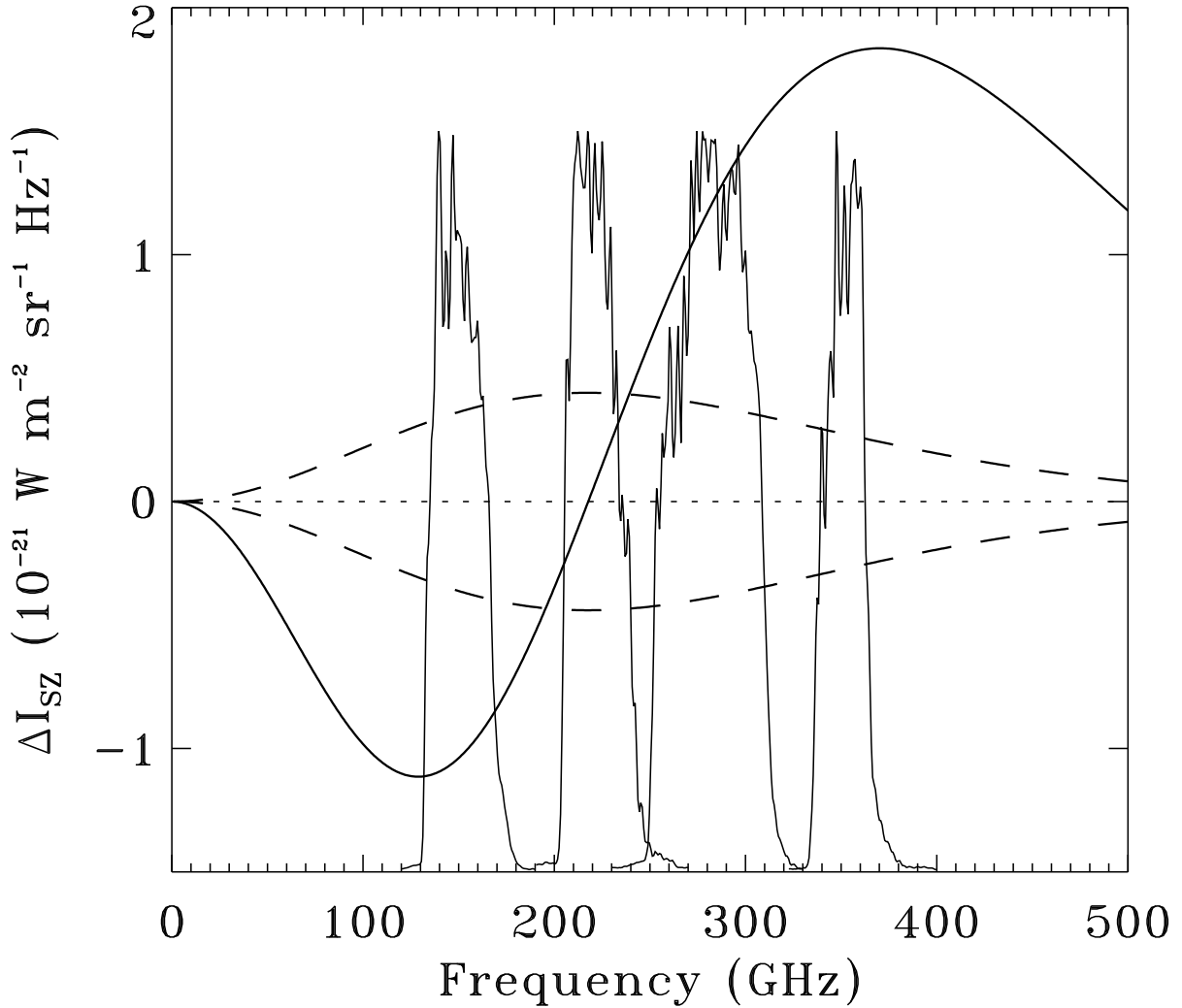}
\caption[SZ Spectra and Measured Bands]{Average {\sc Acbar} spectral bands over-plotted 
on the thermal (solid) and kinetic (dashed) SZ intensity spectra.  The spectral bands
are arbitrarily normalized and have a dynamic range of $\sim-25$dB.  The model cluster 
parameters used are: Compton $y$-parameter of $10^{-4}$, optical depth of $\tau=0.01$,
and peculiar velocity of $\pm1000~{\rm km\,s^{-1}}$.  Refer to \citet{birkinshaw99} for a discussion
of the SZ effects in clusters of galaxies. } 
\label{fig:szbands}
\end{figure}

The {\sc Acbar} frequency bands span the peak of the intensity spectrum of
CMB anisotropies as well as the null of the thermal SZ effect 
(see Figure \ref{fig:szbands}).
Although the atmospheric transmission from the South
Pole is arguably the best on the Earth's surface at millimeter-wave wavelengths, 
care still must be
taken to avoid the deep molecular lines that pervade the atmospheric spectrum. 
We used a model of the atmospheric transmission spectrum at the South Pole in the winter
generated with the AT\footnote{E. Grossman, Airhead Software, Boulder, CO 80302} atmospheric
modeling software.  The frequency bands also avoid sources of astrophysical foreground
emission, such as galactic dust and radio point sources.

The spectral bands were selected by estimating the detector and 
atmospheric noise of the system and
using these to calculate the ratio of astrophysical signal to the quadrature sum of detector and 
atmospheric noise as a function of band center and bandwidth.  
The 150, 280, and 350 GHz
bands were optimized for detection of the SZ thermal effect.  The 220
GHz band was optimized for detection of the SZ kinetic effect (which has the same spectrum
as primary CMB anisotropies).  Because of the constraints placed by atmospheric emission
lines, the 150, 280, and 350 GHz bands are also well optimized for primary CMB fluctuations.
Figure \ref{fig:szbands} shows the spectrum of the thermal and kinetic SZ effects
along with the measured {\sc Acbar} frequency bands.  One can see that the 220 GHz band straddles
the SZ thermal null resulting in very little thermal contamination for measurements of
either the kinetic SZ effect or primary CMB anisotropy.

As described above, the lower edges of the frequency bands are set by the diameter of
waveguide cutoffs in the feed structures.  The upper edges of the bands are defined with
resonant metal-mesh low-pass filters \citep{filters}.  Metal-mesh filters have resonant leaks
at harmonics of the cutoff frequency that must be blocked with additional filtering.  
If left unblocked these high-frequency transmission leaks will allow power from warmer stages 
of the cryostat or hot objects
outside the cryostat to dominate the detector background.

\begin{figure}
\centering
\plotone{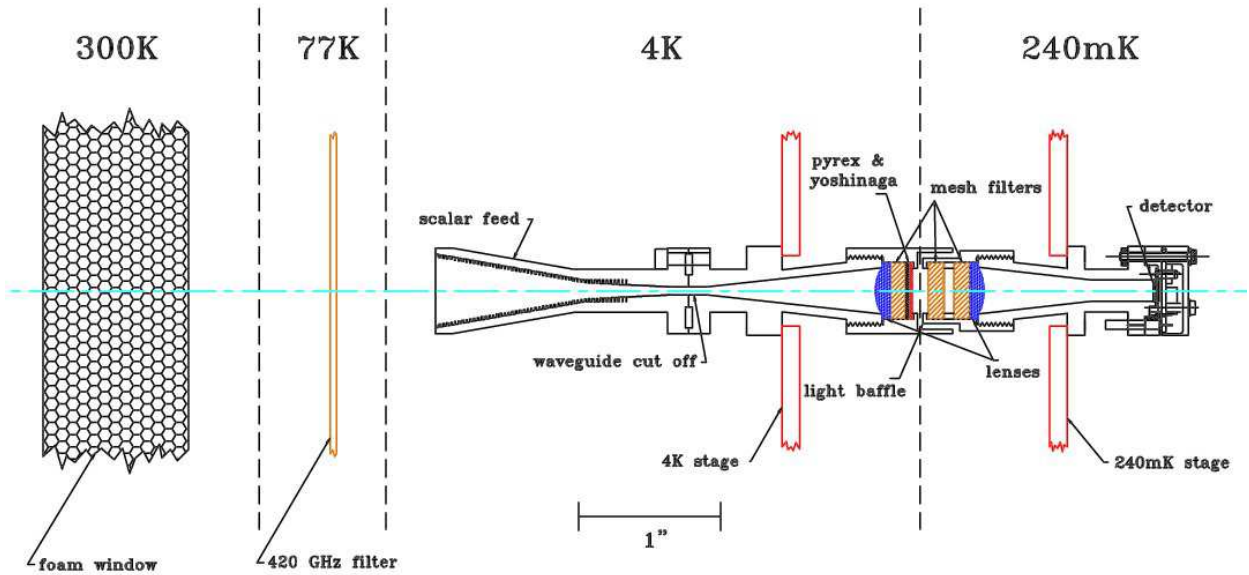}
\caption[Overview of 150 GHz Feed Structure]{{\sc Acbar} 150 GHz feed horn structure and
warm filter arrangement for 2002. 
Dashed vertical lines indicate thermal boundaries.} 
\label{fig:feedstructure}
\end{figure}

High-frequency leaks are particularly insidious because the thermal emission of warm objects
in the Rayleigh-Jeans limit ($h\nu\ll kT$) rises as $\nu^2$ and the 
higher frequencies couple to multiple
spatial modes ($A\Omega/\lambda^2 > 1$).  If these high-frequency leaks are not
blocked at cold temperatures, their cumulative effects can rapidly increase the loading on
the detector as well as couple to undesirable high-frequency sources such as
dust or the atmosphere.  We block high-frequency leaks with a combination of 
additional reflective metal-mesh filters and absorptive Pyrex\footnote{Custom Scientific,
Phoenix, AZ 85020} and alkali-halide (AH)
\citep{yoshinaga} filters.  The band-defining edge filter and one
blocking metal-mesh are mounted at 240 mK, and the third metal-mesh, Pyrex and AH
filters are installed at 4 K (see Figure \ref{fig:feedstructure}).
The filters used in {\sc Acbar} and their cutoff frequencies are listed in Table \ref{tbl:filters}.

All of the filters at 4 K and colder are held in place with threaded filter caps (see
Figure \ref{fig:feedstructure}).  The filters are thermally sunk to their respective feeds
using beryllium copper spring washers inside the filter caps.  Proper heat sinking of the
filters is very important because the low-temperature heat capacity of the focal plane
is dominated by the dielectric materials in the filters.
We also employ a blackened, re-entrant light
baffle across the thermal break between the 4 K feeds and 240 mK feeds.  We use a thin layer
of Stycast\footnote{Emerson $\&$ Cuming, Billerica MA} epoxy mixed with carbon lampblack
as the blackening agent \citep{bock}.  This is applied to the 240 mK
side of the light baffle to reduce optical loading from stray light; the additional optical
loading on the 240 mK stage in negligible.

The set of filters thus described provides sufficient filtering for the optical path, but
the cryostat itself requires filters to reduce thermal loading on the 4 K helium
stage from the 300 K vacuum jacket.  In 2002 we used a single 420 GHz metal-mesh
blocker ($\varnothing$100 mm clear aperture) at 77 K to reduce the load from 300 K (see
Figure \ref{fig:feedstructure}).  In 2001 we had an additional 1.6 THz 
AH filter at 77 K, but we
determined this filter was adding significant optical loading to our detectors; this
filter was replaced with small AH disks within the 4 K section of the feed structure.  

The transmission spectra of {\sc Acbar} were measured with a portable
Fourier transform spectrometer (FTS)
at the South Pole and are shown in Figure \ref{fig:szbands}.  The measured transmission
spectra have been corrected for the $\nu^2$ emission of the FTS source. 
The band centers are determined by
\begin{equation}
\nu_0 = \frac{\int \nu f(\nu) d\nu}{\int f(\nu) d\nu},
\end{equation}
where $f(\nu)$ is the corrected transmission spectrum.
Measurements of bandwidth -- such as the span in frequency at half the maximum value -- 
depend strongly upon the smoothness of the spectrum and its gross
features.  Fringes in a high-resolution spectrum will affect the overall normalization, and
hence, the half-power points.  This is discussed in more detail in Appendix \ref{app:band},
where we present an alternative definition of bandwidth that uses the frequency derivative
of the spectral response as a weighting.
In Table \ref{tbl:boloparams} we present the average
measured band centers and bandwidths of the {\sc Acbar} 2002 
frequency bands.  There were no 350 GHz detectors installed in 2002, so we 
include the optical parameters for 350 GHz from the 2001 observing season in
the table.

The optical efficiencies were determined 
by measuring the optical power absorbed by a detector
while looking into two blackbody loads at different temperatures.  We compare the absorbed
optical power difference with the incident optical power
calculated by convolving the transmission
spectrum with the spectral intensity of the two loads.  
The optical efficiency used to normalize the spectral response, $f(\nu)$, 
of each channel is defined as
\begin{equation}
\eta = \frac{Q_1 - Q_2}{\int f(\nu) [B(\nu,T_1) - B(\nu,T_2)] A(\nu)\Omega(\nu) d\nu},
\end{equation}
where $B(\nu,T)$ is the black body spectral energy density of an object at temperature $T$,
$Q$ is the measured optical power, and $A(\nu)\Omega(\nu)$ is the 
frequency-dependent system throughput 
($\lambda^2$ in the case of a single-moded system).  For this measurement, we used
Eccosorb\footnote{Emerson $\&$ Cuming, Randolph, MA 02368, \#AN-72} microwave absorbing foam 
at both room temperature (300 K) and liquid nitrogen temperature (77 K).  The efficiency
normalized spectral response is thus defined as $\tilde{f}(\nu) = \eta f(\nu)$ and
the optical power absorbed by the 
bolometer from a beam-filling source at temperature $T$ located outside the cryostat is then
$Q = \int \tilde{f}(\nu)B(\nu,T)A(\nu)\Omega(\nu)$.  
The effective optical efficiency over the band was 
calculated by integrating $\tilde{f}(\nu)$ over the band and dividing by the
bandwidth.  This too is discussed in more detail in Appendix \ref{app:band}.  
The effective optical efficiencies for {\sc Acbar}'s 2002
configuration are also given in Table \ref{tbl:boloparams}.

The integrated above-band response (or ``blue leak'') measures the response
to power at frequencies above the
nominal band edge that will couple to undesirable sources.  To measure the out-of-band
response of {\sc Acbar} we used a chopped thermal load and measured the
signal with and without thick-grill filters of varying cutoff frequency.  Thick-grill
filters are plates of metal which have been densely populated 
with cylindrical waveguide holes.  
The filter acts like waveguide passing all light above the waveguide cut off (modulo
the filling factor of the drilled holes).  
We measured the above-band response to a RJ source to be less than 1\% in all channels
in 2001.  The inclusion of additional
blocking filters in 2002 reduced the high-frequency leakage to a level not 
measurable above the noise ($\lesssim 10^{-4}$). 

\subsection{Detectors}\label{ssec:detectors}

{\sc Acbar} detects CMB photons with extremely sensitive micro-mesh spiderweb bolometers
manufactured by the Micro Devices Laboratory at the Jet Propulsion Laboratory (JPL)
and are similar to those
developed for the {\it Planck} satellite \citep{yunspie}.  
These detectors are optimized to detect
broadband emission at millimeter wavelengths.  
The {\sc Acbar} bolometers are background photon noise limited below 300 mK which allows
us to take advantage of the excellent atmospheric conditions of the South Pole.

The {\sc Acbar} bolometers have gold-plated silicon nitride micro-mesh 
absorbers with neutron transmutation doped (NTD) germanium thermistors.  
The spiderweb geometry reduces the cosmic ray cross
section while efficiently coupling to millimeter-wave photons.  The spiderweb
bolometers have very low heat capacity; this results in detector time constants 
of order a few milliseconds.  

The time constants of the detectors were measured on the telescope using a compact, chopped
thermal source mounted behind a hole in the tertiary which we refer to as the ``calibrator.''
The thermal source is an etched metal-film emitter 
manufactured by Boston Electronics\footnote{Brookline, MA 02445, model \#IR-41}.  
The source is small enough that it does not significantly affect the optical loading. 
The chop frequency of the calibrator is
varied from 5 to 200 Hz and we perform a digital lock-in to measure the modulated amplitude at
each frequency.  The signals are then corrected for the transfer function of
the electronics and fit to $\propto (1+i\omega\tau)^{-1}$
to determine the {\it in situ} detector time
constants.  The time constants of the optical bolometers during operation are also listed in
Table \ref{tbl:boloparams}.  The optical time constants of the bolometers are found to be
insensitive to changes in optical loading with elevation.

\begin{figure}
\centering
\plotone{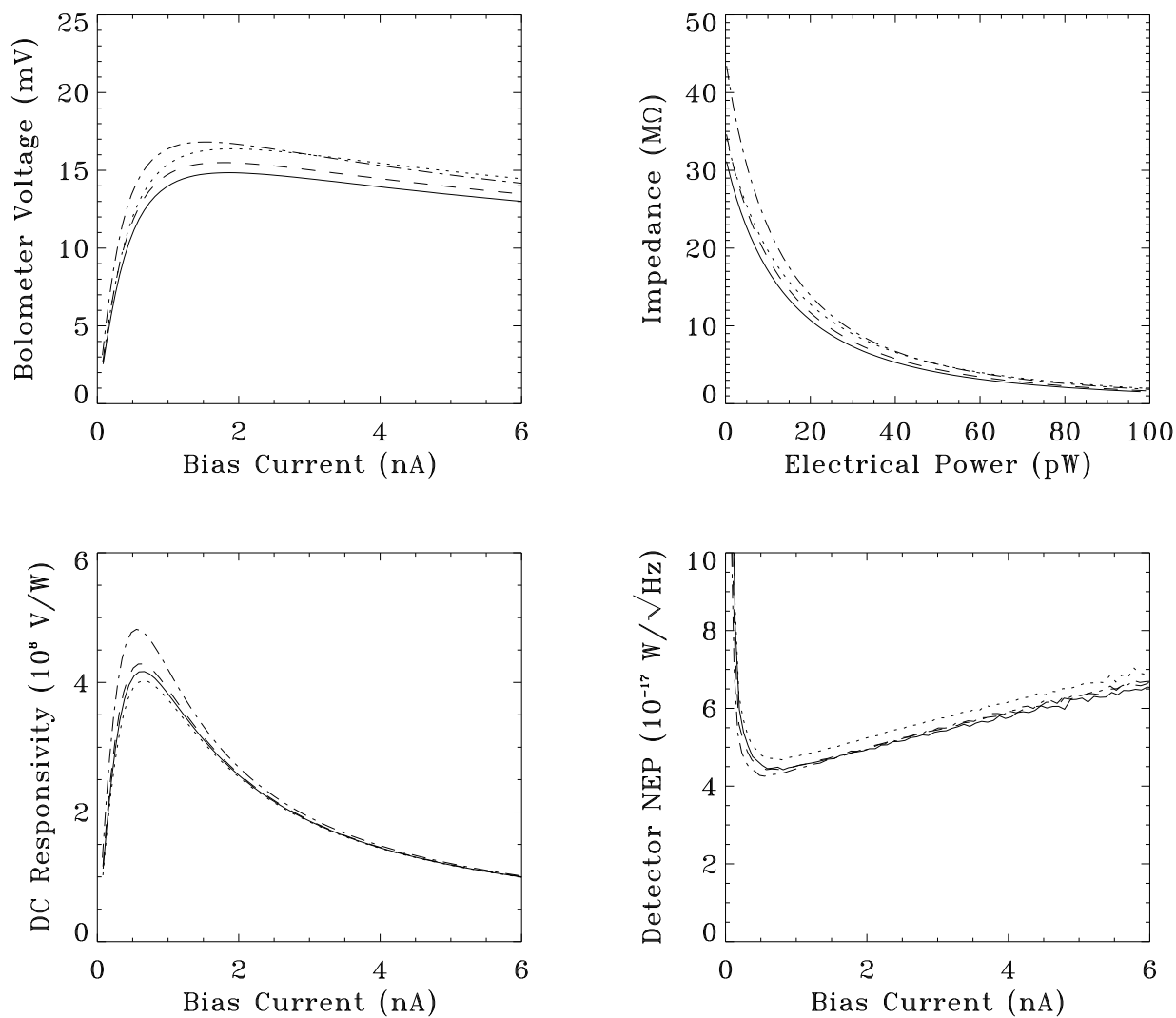}
\caption{Load curves from one row of 150 GHz channels on 2002 March 21.  These 
are four optically-loaded detectors while on the Viper telescope.  The baseplate 
temperature was 238 mK and telescope elevation was $60^\circ$.  
The different line types 
separate the four detectors in the row.  The upper-left panel is the signal 
voltage versus bias current load curve.  The upper-right panel shows the impedance 
of the detectors versus applied electrical power.  The lower-left panel is the DC 
responsivity of the detectors versus bias current.  The lower-right panel is the 
detector $NEP$ (which includes both the Johnson and phonon noise contributions) versus 
bias current.  The detector is shunted by two 30 M$\Omega$ load resistors as 
illustrated in Figure \ref{fig:elect}.} 
\label{fig:lcsb}
\end{figure}

The electrical and thermal properties of the bolometers are measured by using a low-noise DAC to 
slowly ramp the bias current through the detectors and measuring
the output voltage across the detector to produce a load curve.  Analysis of load curves can
provide all of the bolometer parameters of interest ($R_0$, $\Delta$, $G_0$, $\beta$)
as well as the absorbed optical power.  Examples of load curves, responsivity curves, and
detector noise versus bias -- measured with {\sc Acbar} on the telescope in 2002 --
are shown for one row of 150 GHz detectors in Figure \ref{fig:lcsb}.  
The non-ohmic shape of the load curve is due to
the applied electrical power heating the thermistor and causing a decrease in its resistance. 
The bias current applied to the {\sc Acbar} detectors on the telescope is about 2--3 nA
which puts the operating point near the peak of the load curve.  This makes the 
response of the detector insensitive to small changes in
atmospheric loading while achieving near minimum detector noise (as discussed below).  The
measured electrical and thermal bolometer parameters for the {\sc Acbar} detectors 
used in the 2002 season are listed in Table \ref{tbl:boloparams}.

\section{Cryogenics}\label{sec:cryo}

\subsection{Dewar}\label{ssec:dewar}

The {\sc Acbar} dewar (Figure \ref{fig:viper}) is a custom-made 
liquid helium/liquid nitrogen cryostat fabricated by Precision 
Cryogenics\footnote{Indianapolis, IN 46214}.  The
environs at the South Pole are quite harsh with the ambient temperature routinely dropping
below $-100^\circ$F during the austral winter.  To minimize the frequency of cryogen transfers, we
made the cryogen capacity of the dewar as large as would reasonably fit on the telescope
structure.  The outer dimensions of the {\sc Acbar} dewar are $\varnothing20^{\prime\prime}$
in diameter and $36^{\prime\prime}$ in
length (excluding cryogen fill tubes).  The helium and nitrogen tanks each hold 25 liters of
liquid and are wrapped in superinsulation.  
The 4 K cold space is $\varnothing14.15^{\prime\prime}$
by $8.25^{\prime\prime}$ high.  During normal operation,
the liquid helium holds three days including fridge cycles (described below) and the liquid
nitrogen holds about one week.

Because the ambient temperature outside is so cold, we mounted adhesive sheet heaters to the
dewar and electronics boxes, as well as surrounded the instrument with custom made insulating
blankets to keep their temperature above 260 K.  This nominal temperature is still quite
cold and could freeze ordinary rubber o-rings.  We use Ethylene Propylene (EPDM)
o-rings\footnote{Valley Seal Co., Woodland Hills, CA 91367} which are rated to below 250 K
with very low permeability to helium gas.  After nine months of observing in 2001, 
{\sc Acbar} had a mere 15 torr of pressure at room temperature
upon warm-up.  We noticed that the o-rings had permanently deformed after nine months of
observing and replaced them for the second season.

The {\sc Acbar} 4 K radiation shield design employs a re-entrant section to meet the 4 K scalar feed
plate rather than mount a large filter in the top of the 4 K shield with the feeds looking
through it.  This design has two main advantages.  First, only the 16 small waveguide
apertures of the feeds enter the 4 K cold space; these are heavily filtered which greatly
reduces stray optical power in the 4 K space that could load the fridge or bolometers.  The
second benefit comes from the reduction of radio frequency interference (RFI) 
entering the 4 K space by forming a contiguous Faraday shield.

A number of factors influenced the decision to use a foam vacuum window on the dewar rather
than a thin sheet of dielectric (such as Mylar).  Our first concern was scattering of the
beam from thin dielectrics which would result in increased spillover and modulated sidelobe
response.  The dewar window is also quite large ($4^{\prime\prime}$ clear aperture) and we 
were concerned about the strength of the window as well as
its permeability to helium gas.  We measured the transmission, off-axis scattering, and
helium permeability of many materials and selected $1.2^{\prime\prime}$ thick
Zotefoam\footnote{Walton, KY 41094} PPA30 as our window material.  This is a 
nitrogen-expanded polypropylene foam that has very good transmission ($\gtrsim99\%$) 
and low scattering at millimeter wavelengths.  
The closed-cell foam is very strong and has extremely low permeability to helium gas.
We use Stycast epoxy 
to seal the foam to an aluminum mounting ring.  The window deformed permanently when
the dewar was first evacuated, with an inward deviation of $\sim1/4^{\prime\prime}$ 
on the vacuum side across the $4^{\prime\prime}$ aperture.

\subsection{$240\,$mK Refrigerator}\label{ssec:fridge}

The sensitivity and speed of bolometers depends strongly on their operating temperature.  
We want to operate the bolometers at a temperature where the detector noise is well below
the expected photon background limit.  This requires a base temperature below $\sim300$ mK
which is not difficult to achieve with a ${}^3$He sorption fridge.  ${}^3$He fridges have
historically operated from pumped liquid helium baths.  The complication of pumping on the
{\sc Acbar} helium bath while mounted on the telescope was unattractive and we sought an
alternative solution for our detector cooling requirements.
In collaboration with the Polatron \citep{polatron} and Bolocam \citep{glenn} projects, Chase
Research\footnote{ChaseResearch@compuserve.com} 
has developed a three-stage ${}^4$He/${}^3$He/${}^3$He sorption fridge 
that achieves base temperatures
below 240 mK from an unpumped helium bath and was selected for cooling the {\sc Acbar} bolometers.  
The fridge is similar to that described in \citet{bhatia}.

Sorption fridges work on the principle that lowering the pressure above a liquid allows the
molecules with more kinetic energy to escape, thus cooling the liquid until the vapor
pressure equals the pressure above the liquid.  This is the same principle that makes
cooking pasta at higher elevations difficult because water boils at a substantially lower
temperature.  This effect has an added benefit for {\sc Acbar} at the Pole because the low
pressure at $9,300^\prime$
elevation pumps the liquid helium bath of the cryostat
and drops its temperature to $\sim3.9$ K instead of
the 4.2 K at sea level pressure.  This small change in base plate temperature greatly
improves the condensation efficiency of the fridge cycle; the hold time of
the fridge at the Pole doubled from the hold time achieved in 
Berkeley during system integration.  During
normal operation, the intercooler stage cools to 370 mK for approximately 32 hours
and the ultracold stage operates near 240 mK for a longer duration.  The fridge is 
automatically recycled
when the intercooler is exhausted and takes about three hours from the start 
of the cycle to below 250 mK.  The duty cycle of the fridge is $\sim90\%$.

The bolometer baseplate temperature provided by the ultracold still is very stable
over the course of a 5 hour CMB observation with an RMS scatter of $<100~\mu$K.  
Because {\sc Acbar} is mounted directly on
the telescope, we were concerned about variations in baseplate temperature with changes in
telescope elevation.  
The refrigerator was constructed with copper sinter in the helium stills to improve
the coupling of the liquid to the metal housing.  The baseplate temperature
is measured as a function of zenith angle during the course of a skydip and that there is
$<1$ mK change over a range of 75$^\circ$ in dewar angle.

\subsection{Thermometry and Control}\label{ssec:thermo}

{\sc Acbar} employs three different types of thermometry inside the cryostat; all 
of these are
manufactured by Lake Shore Cryotronics, Inc\footnote{Westerville, OH 43082}.  
For temperatures between $\sim4$ K and 300 K we use two-wire silicon diode 
thermometers.  
These are used on both liquid cryogen tanks, the JFET module, all three 
fridge pumps, and the two heat switches.  
On the cold stills of the fridge, we use calibrated four-wire germanium
resistance thermometers (GRT) which are useful for temperatures below 4 K.  
The ultracold bolometer stage is equipped with a calibrated four-wire Cernox 
RTD which is only accurate to $\sim5$ mK but has a dynamic range from below 
$230\,mK$ to a few hundred Kelvin.  This Cernox thermometer is particularly 
useful because we can monitor the temperature of the bolometer stage during 
cool down as well as measure its temperature during normal operation with the
same sensor.
In addition to these sensors, the dewar has a level monitor installed in
the helium tank.
A current is briefly applied to a superconducting strip immersed in the 
helium tank; the sensor quickly becomes normal over the portion that is 
not immersed in liquid helium.
Periodic monitoring of the helium level is performed by measuring the resistance 
of the sensor.
The helium level monitor has proved invaluable to monitoring helium transfers 
in the harsh environment at the South Pole. 

The refrigerator cycle involves a sequence in which 
the pump heaters and gas-gap heat switches are periodically activated.
An electronics box mounted directly on the cryostat contains 
8 programmable heaters, 12 diode sensor readouts, and 4 auto-ranging GRT/Cernox
readouts, and the helium level sensor readout.
A computer program monitors the relevant temperatures in the system and 
orchestrates the application of power to control the pumps and heat 
switches.

\subsection{Thermal Isolation and Heat Sinking}\label{ssec:iso}

The limited cooling capacity of the fridge requires us to restrict
the thermal loading on the
bolometer stage to $<3~\mu$W of power to keep the temperature below 250 mK.  
The challenge was to rigidly support the
massive copper focal plane as well as read out all of the bolometer signals 
and associated thermometry while keeping the thermal load on the fridge to a 
few $\mu$W.

The scalar feeds are mounted on a gold-plated aluminum plate which is rigidly 
fixed to the 4 K helium cold plate via $\varnothing 0.5^{\prime\prime}$ 
gold-plated copper rods.  The structural support
of the cold focal plane is provided via Vespel\footnote{DuPont Engineering 
Polymers, Newark,
DE 19714-6100} stand-offs and a tensioned Kevlar harness. 
The Vespel stand-offs are machined to $\varnothing 0.25^{\prime\prime}$
and $0.015^{\prime\prime}$ wall-thickness and have threaded aluminum
end caps.  The tensioned Kevlar reduces mechanical vibrations in the focal
plane which can cause microphonic pickup 
(discussed in more detail below in \S\ref{ssec:sigelect}).

\begin{figure}
\centering
\plotone{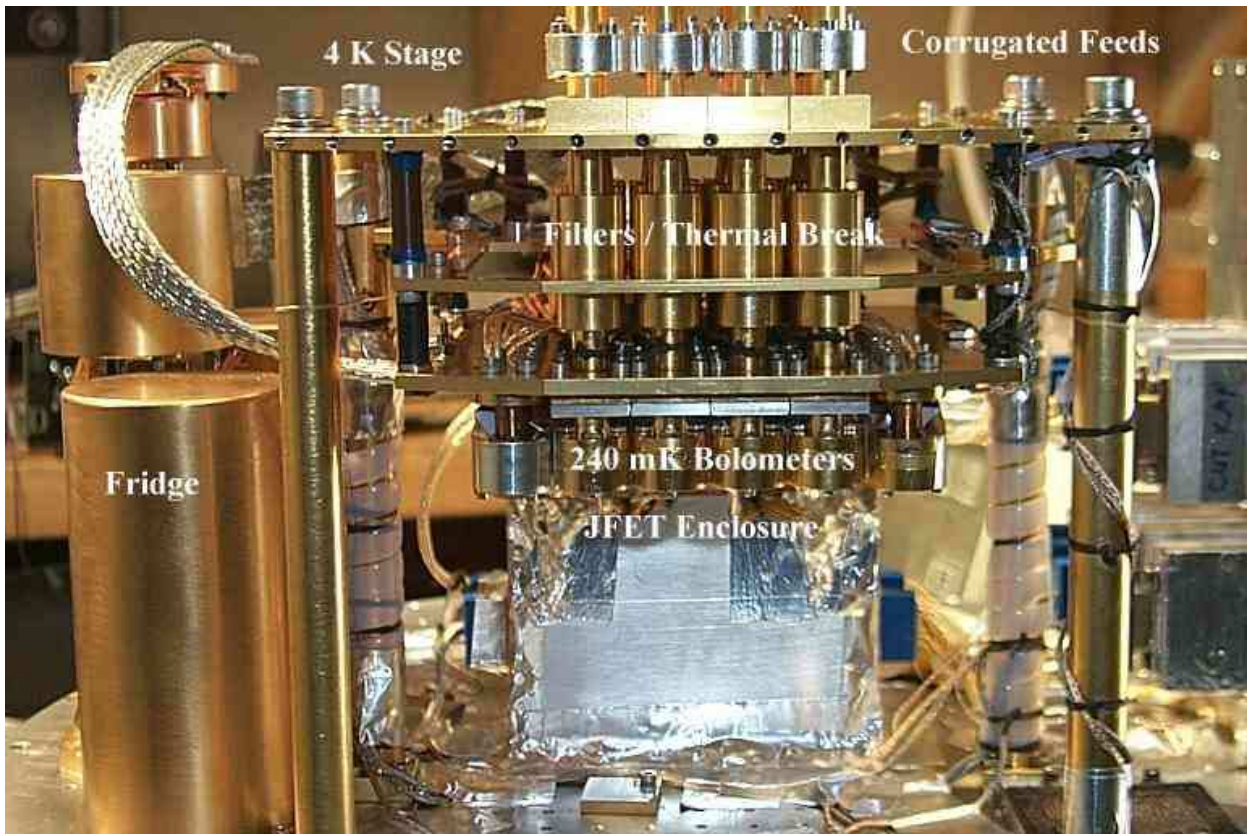}
\caption[Photo of Focal Plane Structure]{Photo showing the different elements
of the {\sc Acbar} focal plane.} 
\label{fig:fpfront}
\end{figure}

We use the $\sim 100~\mu$W of cooling power of the 370 mK intercooler 
still as
a thermal buffer between the 4 K horn plate and the 240 mK detector stage.  
This intercepts most of the heat load that would otherwise overwhelm the 
small heat lift of the ultracold stage.  
There are multiple types of Vespel material and 
we use SP-1 between 4 K and 370 mK and SP-22 between 370 mK and 240 mK.  
We estimate the thermal loads from the Vespel legs to be $\sim40~\mu$W
on the intercooler stage (370 mK) and $\lesssim 0.1~\mu$W on the ultracold 
stage (240 mK).  
Figure \ref{fig:fpfront} shows an image of the {\sc Acbar} focal plane structure 
along with the sorption fridge.

Flexible thermal straps (see Figure \ref{fig:fpfront}) are used to couple 
the fridge stills to the two thermal stages.  These are made from 
nickel-plated copper braid in which indium
is embedded at each end for clamping. The flexible thermal straps reduce 
the coupling of mechanical resonances of the fridge to the focal plane while 
providing good thermal conductivity.

All of the isothermal wiring inside {\sc Acbar} is Teflon-coated gold-plated 
copper wire and all
of the wiring that traverses multiple temperature regions is Teflon-coated 
stainless wire.  
Both of these are surgical-quality wires manufactured by Cooner 
Wire\footnote{Chatsworth, CA 91311}.  The stainless wiring is bundled 
into six twisted pairs within a common stainless
shield (corresponding to the six detector channels per bias) with no 
insulating outer jacket.  
Stainless wiring has low thermal conductivity but relatively high 
electrical impedance. All
of the wiring for high-current devices (such as resistive heaters) is 
doubled-up to reduce
the power dissipation in the wiring.  The wires joining stages at different
temperatures are heat sunk at each stage by embedding the wires in a 
gold-plated copper tab with Stycast epoxy.  The stainless wiring
is estimated to contribute less than 50 $\mu$W of power on the 
intercooler stage and around 0.5 $\mu$W on the ultracold stage.

\section{Electronics}\label{sec:electronics}

\subsection{Signal Electronics}\label{ssec:sigelect}

\begin{figure}
\centering
\plotone{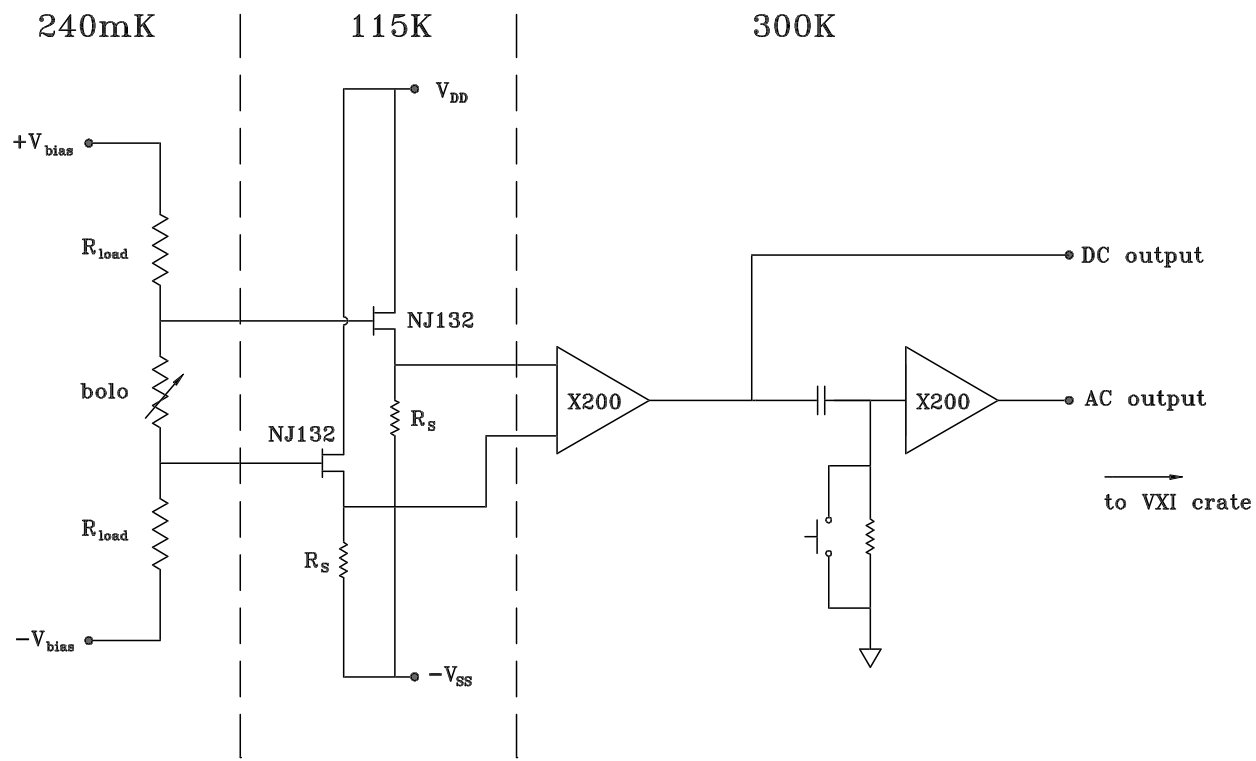}
\caption[Signal Electronics]{Schematic of the {\sc Acbar} signal electronics chain. 
Dashed vertical lines denote thermal boundaries.} 
\label{fig:elect}
\end{figure}

The {\sc Acbar} signal electronics are designed to provide DC bias across the bolometers and read
out all channels with low noise from nearly DC to $\sim100$ Hz.  Figure
\ref{fig:elect} shows a schematic of the {\sc Acbar} signal electronics chain.  The DC bias board
supplies a low-noise voltage of 0 to 500 mV symmetrically across the bias resistor
stack.  There are four bias voltages -- one for each row of detectors -- which can be set
independently and are applied to six detectors each.  Two sets of twisted pair stainless
wires (one set for redundancy) bring each of the bias voltages to the focal plane where they
are broken out to six detectors.  
In addition to the four optically-loaded bolometers per bias, each of the 
four biases is also applied to a
``dark bolometer" (a bolometer which has been blanked off with a blackened load at 240 mK)
and a ``fake bolometer" (a 10 M$\Omega$ 
resistor in place of the bolometer) for use as monitors of base
plate thermal drifts, microphonic noise, and amplifier noise.

Each bolometer is in series with two 30 M$\Omega$ metal-film load resistors
which were custom made by Mini-systems, Inc.\footnote{Attleboro, MA 02703}.  The load
resistor package is surface mounted to a PCB board which is epoxied directly onto the back of
the bolometer module.  Also mounted on the PCB board is an RFI filter on each side
of the bolometer composed of surface-mount 47 nH inductors and 10 pF capacitive feed-through
filters\footnote{muRata part $\#$'s LQP21A47NG14 and NFM839R02C100R470} which provide
filtering above a few hundred MHz directly on the bolometer module.

\begin{figure}
\centering
\plotone{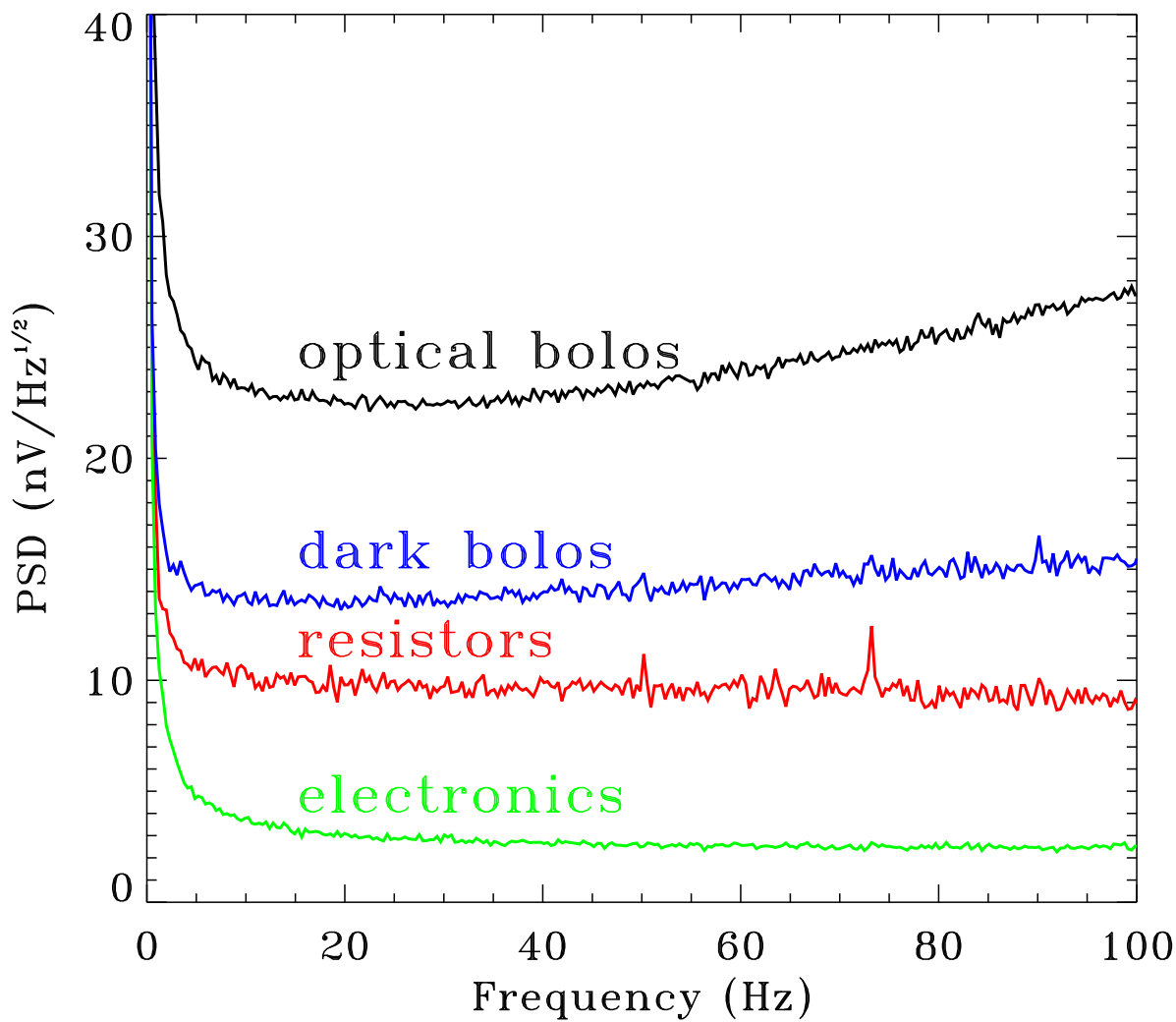}
\caption[Electronics Noise]{Average voltage noise PSDs of the {\sc Acbar} signal channels in 2001.
The electronics noise (green/bottom) is measured 
from a pair of JFETs with their gates shorted together and 
shunted to ground with a 10 k$\Omega$ resistor.  The resistor channel (red/second) is for a 10 
M$\Omega$ ``fake bolometer'' mounted on the 240 mK stage.  
The dark bolometers (blue/third) are blanked off at 240 mK
and the optical bolometers (black/top) are the average of all 16 optical channels.  A narrow
60 Hz line
filter was applied and we corrected for the transfer function of the electronics.
As shown in Figure \ref{fig:net150}, the used signal band is less than 60 Hz
and frequencies above the band are removed with a step-function filter.
The rise in the optical bolometer noise at high frequency results from correcting
for the detector time constants.  This is because the Johnson and amplifier noise 
contributions are not attenuated at high frequency by the detector time constants.
}
\label{fig:vnoise}
\end{figure}

The high impedance ($\sim 7\,{\rm M}\Omega$) of the bolometer signals makes them 
susceptible to microphonics caused by vibrating wires and RF pickup. 
In order to reduce these potential sources of noise, pairs of matched JFETs 
are placed close to the detectors to reduce the output impedance of the signals. 
The bolometer signal voltages are sent to the JFET modules on twisted pairs of stainless 
wire.  
The bolometer signals of a common bias are carried by a shielded bundle of six twisted
pairs.  
Each side of the bolometer voltage is sent into one side of a matched NJ132 JFET
source follower pair (see Figure \ref{fig:elect}). 
The JFET dies were manufactured by Interfet\footnote{Garland, TX 75042};
they are wire bonded to a single alumina substrate that holds 24 pairs of 
matched JFETS.
These devices have low noise ($\sim 2~{\rm nV\,Hz^{-1/2}}$ and an extremely 
low 1/$f$ knee ($\sim 1~{\rm Hz}$); the noise from the JFET followers is small
compared to the contributions of the detectors and atmosphere (refer to Figure
\ref{fig:vnoise}).
The dies were probed and selected to form pairs with $<20\,$mV difference in 
gate-source voltage; this criteria guarantees that the sum of the offset and 
the detector signal will not saturate the analog to digital converter 
after applying a gain of 200.

The JFETs cannot operate at 4 K and achieve their lowest noise performance at 
a temperature of $\sim 115\,$K.
The entire JFET module is stood off from the 4 K cold plate with low thermal conductivity
G-10 fiberglass legs. A heater is used to raise the temperature of the JFETs to 
their operating point.
In order to minimize the heat load on the helium stage, the heat from the JFET
module is 
intercepted half way down the G-10 legs by a 77 K cold finger that extends from the 
nitrogen can through the helium can (see Figure \ref{fig:viper}).

In order to reduce vibrations of the high impedance wiring connecting the detectors
and JFETs, the JFET module and detector array are both secured with a tensioned Kevlar 
support harness.
The Kevlar strands are tensioned with spring washers to overcome the negative coefficient 
of thermal expansion of Kevlar.  
The addition of this harness pushes the vibrational resonance frequencies of the
focal plane and JFET module well above the signal band.  

RFI entering the 4 K vacuum space can couple to the high
impedance wiring on the focal plane; this can heat the detectors and degrade their
sensitivity or introduce noise to the signals.
All wires entering the 4 K vacuum space pass through
additional RFI filter modules mounted in the wall of the liquid helium radiation shield.  
The filters used in these modules are muRata EMI $\pi$-filters\footnote{muRata part
$\#$VFM41R01C222N16-27} which have been embedded, along with the wires from the 
connectors, in castable Eccosorb\footnote{Emerson \& Cuming part \#CR-124}.
These RFI filter modules were measured to attenuate signals above 1 GHz better than
-60 dB \citep{leong}.  

The signal wires exit the dewar through hermetic connectors
and enter the warm electronics signal box where they
are filtered and amplified.  The electronics boxes are mounted directly
on the cryostat.  The boxes have RF shielding gaskets at all mating
surfaces.  
The raw signals are low-pass filtered at $650\,$Hz and amplified by a factor 
of 200.  
The DC offset is removed with a low-pass filter at $\sim16$ mHz 
($10\,$s time constant).  
In order to avoid a long wait for the readout to settle after a change in
elevation, a MOSFET switch is used to momentarily short the capacitor in 
the low-pass filter and remove the DC offset. 
The AC signals are amplified by another factor of 200 (for a total gain 
of 40,000).  
The total noise contribution from the JFETs and warm electronics is 
$<10~{\rm nV\,Hz^{-1/2}}$ for $\nu > 2$ Hz with a white noise level of 
$\sim 3~{\rm nV\,Hz^{-1/2}}$ for $\nu > 10$ Hz.  Figure \ref{fig:vnoise}
shows the voltage noise of the electronics, resistor channels, 
dark bolometers, and optically loaded bolometers.  The electronics noise
contributes a small fraction of the total noise of the optically loaded
detectors over most of the signal bandwidth.

The signals are then low-pass filtered at 650 Hz and
sent to an Agilent Technologies VXI data acquisition system housed in
a heated enclosure on the back of the telescope primary mirror.  There the signals are
anti-alias filtered by a 1-pole filter with 3dB point at $100\,$Hz, 
sampled with 16-bit resolution at $2400\,$Hz, and 
averaged over 8 samples to reduce the data rate.  The averaging
puts the Nyquist frequency at 150 Hz and does not effect our signal bandwidth.  
We transfer the data to the U.S. using the TDRSS satellite network.  This allows the
local science team to monitor the state of the instrument and analyze data within a day of
acquisition.

\subsection{Transfer Function}

Modulated optical signals undergo a considerable amount of filtering
between absorption by the detectors and the digitized voltages written to disk.
The transfer function of the system quantifies the amplitude attenuation and
phase shift of modulated optical signals as a function of modulation frequency.  
The filter
components of the {\sc Acbar} transfer function are single-pole detector time constants,
$RC$ filter from detector impedance and capacitance from wiring and JFETs, 650 Hz
low-pass $RC$ filter, 100 Hz single-pole anti-aliasing filter in the A/D, and
a sinc filter from the box-car averaging of 8 samples.  We measure the complete
transfer function by varying the frequency of a small, chopped, thermal source 
from 5 to 200 Hz and record the amplitude and phase of the output signal. 
The time-stream data for each optical channel is corrected for the 
complete transfer function with the appropriate detector time constants.

\subsection{Computer Control and Housekeeping}\label{ssec:housekeep}

{\sc Acbar} can be configured for either manual or computer control of all system elements by
flipping large toggle switches (for ease of use with gloved hands) on the electronics
boxes.  
Under manual control, all receiver settings can be made with turn pots.  
Under computer control the settings are changed with a digital bus.  
The digital bus allows remote setting of all of the following: bolometer
bias levels, all heaters on fridge and cold stages, calibrator temperature and modulation
frequency, and all thermometer settings (reference impedance and excitation voltage).  This
allows the observer to control virtually all aspects of the instrument from within the
main station dome, located approximately 1 km away from the telescope.  All of the
housekeeping information is read by the A/D at 2 Hz and saved to disk.  This includes
all bias levels, thermometry readings, DC levels of all 24 detectors, and the two-axis tilt
meter mounted on the telescope.

\section{Observations}\label{sec:obs}

\subsection{Beam Sizes}\label{sec:beams}

Accurate measurement of the beam sizes on the sky is important for calibrating the instrument
and generating CMB window functions.  
The planets Mars and Venus are our objects of choice for
measuring beam sizes.  The beam solid angle is determined by integrating the source voltage
over raster maps (that have been flat-fielded) and dividing by the source amplitude.
The measured beam sizes from Mars on 2001 July 17 and Venus on 2002 September 22 are listed in 
Table \ref{tbl:beams}.

The beam sizes for the 150 and 350 GHz channels differ from the nominal $4^\prime$
Gaussian FWHM. 
The 150 GHz beams are diffracted by the aggressive edge taper (-18dB) on
the 2 m primary mirror and the $\sim20\%$ increase in size agrees with the level predicted
for a truncated Gaussian \citep{goldsmith}.  
The 350 GHz beams are properly sized exiting the cryostat and are not improved by
focusing the telescope; the large beams sizes are most 
likely the result of telescope surface roughness.

The beam size varies within a frequency column/row, 
indicating some aberration across the focal surface.  
We chose the dewar focus position that, on average, resulted in
the minimum beam size from galactic source raster maps.  We made a particular 
effort to minimize the beam size of the 150 GHz channels because they are the 
most sensitive for CMB and SZ observations.

The beam sizes change slightly with chopper angle, indicating 
that the beams are distorted as the chopper rotation changes the optical path 
through the telescope.  The effect was measured to be $\sim5\%$ in 2001 but was
reduced by about a factor two in 2002 with an improved alignment of the optics.   The
changing beam sizes also affect the CMB window functions of the instrument.  This
effect is discussed in detail in \citet{kuo02}.
The conservation of optical throughput with chopper angle
was tested by making complete raster maps of galactic sources at
multiple chopper offset positions and integrating the maps over solid angle. 
We used the compact HII region MAT6a \citep{puchalla} 
for this measurement because of the limited
availability of planets and bright point sources.
The integrated source voltage of MAT6a versus chopper position
is found to be flat, indicating good conservation of optical throughput.

In late September 2002 we noticed that ice crystals had accumulated on the foam dewar window
and removed them.  From maps of Venus before and after the ice removal, we discovered a
significant change in some of the beams.  The ice on the curved surface of the window acted
like a thin lens, distorting the focal plane.  
The average FWHM beam sizes before the ice removal were
$4.89^\prime$, $4.30^\prime$, and $4.23^\prime$ at 150, 220, and 280 GHz, respectively.  
After the ice removal, the average beams measured $4.69^\prime$, $4.25^\prime$, and
$3.95^\prime$.  In addition, the positions of some of the beams were
shifted by a substantial fraction of an arcminute on the sky.  
Both of these effects tend to smear the effective
beam size of the coadded maps.
Fortunately, we included a bright point source in our CMB fields (see \S\ref{sec:fields})
and observed galactic sources multiple times per day; these allow us to measure
changes in the position and size of each beam and correct for the effects of ice accumulation
over time.

\subsection{Chopper Synchronous Offsets}\label{sec:offsets}

When the chopper is running, the bolometer signals are dominated by 
a chopper synchronous offset that is well described by a second-order polynomial.  
The chop across the sky
is roughly constant latitude at $45^\circ$ elevation and we would expect an offset 
from motion of the beams through the atmosphere at different elevations.
The amplitude of the temperature
offset due to motion through the atmosphere is approximately equal to
\begin{equation}
\Delta T \sim 2.6~{\rm mK}\left({\frac{T_{sky}}{220~{\rm K}}}\right)
\left({\frac{\tau}{0.03}}\right) \left({\frac{\Delta}{1^\prime}}\right)
\end{equation}
where $T_{sky}$ is the temperature of the atmosphere, $\tau$ is the atmospheric opacity,
and $\Delta$ is the elevation change from the curvature of the chop.  
However, the observed temperature
change is at times much larger than the predicted amplitude from
the atmosphere, 
and sometimes has the opposite sign.
This indicates the offset is often dominated by effects other 
than the changing elevation of 
the beams through a smooth atmosphere.

The offset structure is observed to depend on a number of factors.  Modulated chopper
spillover,  accumulation of snow on the telescope, and atmospheric 
conditions appear to be the dominant contributors to offset amplitude.  
The first two sources are within our ability
to control and we take steps to mitigate their effects.  We reduced modulated mirror
spillover by mounting a blackened circular light baffle between the tertiary mirror and the
chopper.  This light baffle is somewhat smaller than the projected area of the chopper and
intercepts beam power which would otherwise spill over and modulate as the chopper rotates.  
The baffle significantly reduced the offset amplitude of the chopper offset.  As discussed
below in \S\ref{sec:loading}, this warm baffle contributes to the loading of the
system because it intercepts beam power at ambient temperature.  

\subsection{Snow}\label{sec:snow}

Snow accumulation is another large contributor to the chopper synchronous offsets.  The
mirror surfaces collect blowing snow as well as develop a thin layer of frost
over time.  As the chopper rotates,
the detector beams view different projections of the chopper mirror and are swept 
across the secondary mirror.  The chopper is placed at an image of the primary, so 
the illumination of the primary mirror is approximately constant.  
Unevenly distributed snow on the chopper and secondary mirrors 
contributes an optical signal (or ``chopper offset") through emission and 
scattering.  To avoid this, we clean and defrost the mirrors frequently.

In addition to contributing to offset structure, snow accumulation also attenuates
astronomical signals.  
The method of signal loss is likely to be a combination of
absorption and scattering by
crystals with size comparable to $\lambda$.  This signal attenuation was discovered by
comparing the measured amplitude of galactic sources versus chopper offset amplitude.  To
test the effects of snow on signal attenuation we performed raster maps of the 
galactic source RCW38 with the
mirrors very snowy and then repeated the observation with the mirrors cleaned.  The average
signal ratios for RCW38 with the mirrors snowy versus clean were 70\%, 45\%, and 20\% at
150, 220, and 280 GHz, respectively, indicating a strong frequency dependence of the
signal loss.  The beam
solid angles are effectively unchanged between the snowy versus clean measurements,
which indicates that any scattering would have to be relatively isotropic to explain 
the signal loss without an appreciable broadening of the beams.  
We minimize the effects of snow accumulation by identifying periods when the 
data may be contaminated from snow and discarding them.

To develop a criteria for cutting data with snow contamination, we investigated the
correlation of chopper offset amplitude with integrated
source signal and found a strong correlation at
high frequency and a weaker correlation at low frequency.
This frequency dependence allows us to tune the level of tolerable
snow attenuation to the science goal being investigated.  For example, the weaker dependence
at 150 GHz means that a more aggressive 
snow cut threshold (fewer files removed) can be used for science only
incorporating the low-frequency channels.  However,
pointed cluster observations rely on the high-frequency data to remove the CMB
contribution from the maps and must therefore employ a more conservative snow cut threshold
(more files removed) to prevent severe signal attenuation at the higher frequencies.  

\subsection{Optical Loading}\label{sec:loading}

The optical loading on all channels can be determined at any time from the DC 
voltage levels of the bolometers and the detector cold plate temperature using the 
measured properties of the bolometers. The average RJ loading temperature 
for 2002 at all three frequencies is listed in Table \ref{tbl:noise}.  The
variation in optical loading is found to correlate well with 
atmospheric opacity.

The dominant sources of loading for {\sc Acbar} are warm filter optics, atmosphere, and a warm
telescope.  We took great pains to reduce the internal optical loading from the dewar by
maximizing the in-band transmission of the warm filters and blocking high-frequency leaks at
cold temperatures.  The atmospheric contribution is given by $T_{atm} = T_{sky}
(1-e^{-\tau/\cos\zeta})$ where $T_{sky}$ is the physical temperature of 
the atmosphere, $\tau$ is the
average in-band zenith opacity, and $\zeta$ is the zenith angle of the observation.
On a typical day, the temperature loading from the atmosphere at $EL=60^\circ$
is 8, 13, and 24 K at 150, 220 and 280 GHz, respectively.
The telescope consists of four warm aluminum mirrors, each of which has a
theoretical emissivity of a fraction of a percent from Ohmic loss.
The actual emissivity of the surface is greater than this
value because of reduced conductivity from surface roughness.  Surface roughness
also contributes to loading through Ruze scattering onto warm surfaces.
The introduction of a light
baffle to reduce chopper synchronous offsets also contributes to the optical loading.  The
large beams of the 150 GHz channels are truncated at the baffle aperture at a level which
contributes $\sim5$ K, but the beams from the higher frequencies are sufficiently small that
the baffle contribution is $\lesssim 1$ K.  The average total RJ loading on the bolometers
at $60^\circ$ elevation in 2002 (and associated $1\sigma$ dispersion) 
was $39\pm6$ K at 150 GHz, $37\pm10$ K at 220 GHz, and $65\pm15$ K at 280 GHz.  The variation 
in loading is highly correlated with changes in atmospheric opacity.

\subsection{Scan Strategy}\label{sec:scanstrat}

\begin{figure}
\centering
\plotone{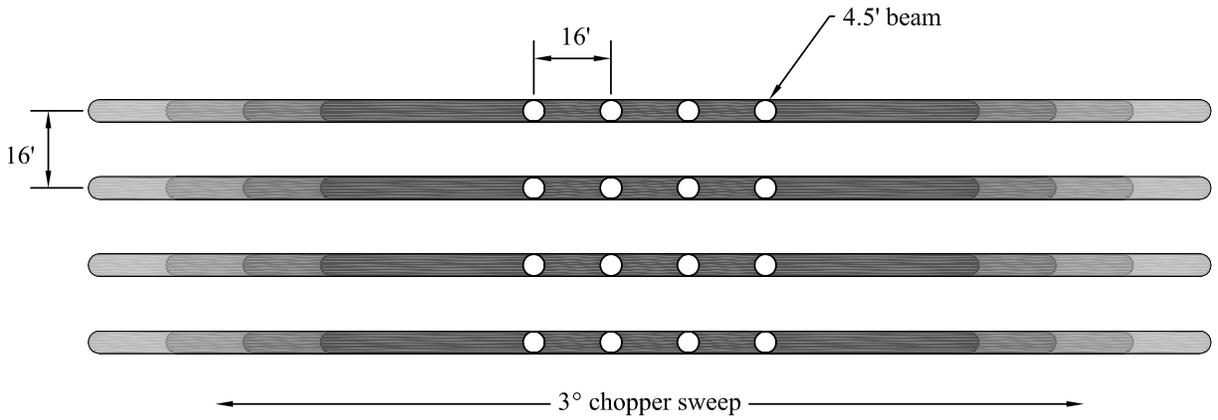}
\caption[Chop Sky Coverage]{Illustrates the sky coverage geometry of a single declination 
``stare.''  The 16 pixels are swept $\sim3^\circ$ on the sky and the darkness 
of the shading indicates 
the degree of overlap of the four channels within each row.  Full raster maps are built up by 
tipping the telescope down $1^\prime$ in elevation and repeating this process approximately 
100 times.} 
\label{fig:chop}
\end{figure}

We perform elevation raster maps when observing the CMB with {\sc Acbar}.  
We track a fixed RA-DEC position with the chopper running for 30 to 60 seconds; this is
referred to as a ``stare.''  We then tip the
telescope down $1^\prime$ in elevation and repeat the process.  
We perform $\sim100$
elevation steps which results in a large patch of sky sampled by all four rows of 
the focal plane.  
The sky coverage of the array during a single stare is illustrated in Figure
\ref{fig:chop}.  

As described above, the dominant signal in the raw raster maps is a roughly
parabolic signal a few mK in amplitude.  The chopper synchronous offset signals change with time
and we were concerned that changing small-scale variations in the offset signal could
contaminate the maps.  We devised an observing strategy to remove as much chopper
synchronous offset as possible -- even if slowly time varying -- while preserving the 
large-scale CMB power in the map.

We employ a LEAD-MAIN-TRAIL (LMT) observing strategy in which we raster three
$\sim3^\circ\sec\delta$ wide fields that almost touch in RA.
The raster progresses by
observing the three fields in succession at fixed elevation before proceeding to the next
elevation.  The fields are usually observed for 30 seconds on LEAD, 60 seconds on MAIN, and
30 seconds on TRAIL.  If the offset is constant or
changing linearly with time, then the average offset
of the LEAD and TRAIL fields should equal the offset in MAIN.  By forming the difference
$LMT = M-(L+T)/2$, we eliminate both constant and linearly drifting offsets.
We also divide the observations into $A$ and $B$ halves which are
shifted by $\sim0.5^\circ\sec\delta$ in RA.  This shift provides an additional test
of systemic effects.

The three fields are separated by about $3^\circ$ on the sky, which is larger than the
$\sim1^\circ$ peak correlation of the CMB;  thus the $LMT$ subtraction should not remove much CMB
power.  In fact, if the CMB fluctuations are random and uncorrelated, then the $LMT$ map
should have $\sqrt{3/2}$ times the CMB signal of the MAIN field alone.  
When searching for SZ clusters, we are not concerned with preserving the large-scale CMB
power in the maps.  We can treat the three $LMT$ fields as separate and remove an average offset
plus higher order polynomial to eliminate the chopper synchronous signal;
this preserves the unique frequency spectrum of the thermal SZ effect so that
clusters will always appear as decrements at 150 GHz in the
three separate fields.  This is discussed in more detail in \citet{runyan03b}.

\subsection{Field Selection}\label{sec:fields}

When selecting fields for deep CMB observations at millimeter wavelengths, the primary
foreground contaminant of concern is dust emission \citep{tegmark96}.  Most of the
Southern celestial hemisphere is available for continuous observation and 
we select regions of very low dust contrast for our observations.  
The IRAS/DIRBE dust map of \cite{finkbeiner99} 
provides a template of galactic dust emission with the
best region of the Southern Hemisphere lying roughly between $21^h$ to $5^h$ in RA 
and $-20^\circ$ to $-70^\circ$ in DEC.

With thousands of square degrees of clean sky available, 
we decided to center the fields on millimeter-bright point sources.
These sources provide a continuous monitor of pointing during CMB observations, as
well as a bright point source for measuring the final beam size in the coadded maps.  The
coadded point source image incorporates the physical extent of the beams as well as beam
smearing due to pointing jitter.  We surveyed many of the flat-spectrum radio sources in
the Southern Hemisphere observed by SEST \citep{beasley, tornikoski96, tornikoski00}
searching for candidates bright enough to detect with high S/N in a single raster map. 
The sources selected for the CMB fields observed by {\sc Acbar} in 2001 and 2002 are listed
in Table \ref{tbl:quasars}.

\subsection{Data Cuts}\label{ssec:cuts}

We apply a variety of cuts to the raw data set before inclusion in
the coadded maps.  Our total observing efficiency during the winter of 2002,
including galactic source observations and calibrations, was $\sim60\%$
before any data cuts were applied.
The first data cut is based on the reliability of the pointing solution for a
given observation.  There are brief periods of time where the observation of galactic
sources does not yield a consistent pointing solution and observations during these
periods are excluded.  
We next verify that the $^3$He refrigerator base temperature is cold ($<$ 250 mK) during
the entire observation to prevent significant changes in detector responsivity; if the
refrigerator is still cooling down or warming up during an observation, we do not include 
that data.  We also cut an entire chopper sweep for a channel with a detectable cosmic
ray spike above $5\sigma$.

\begin{figure}
\centering
\plotone{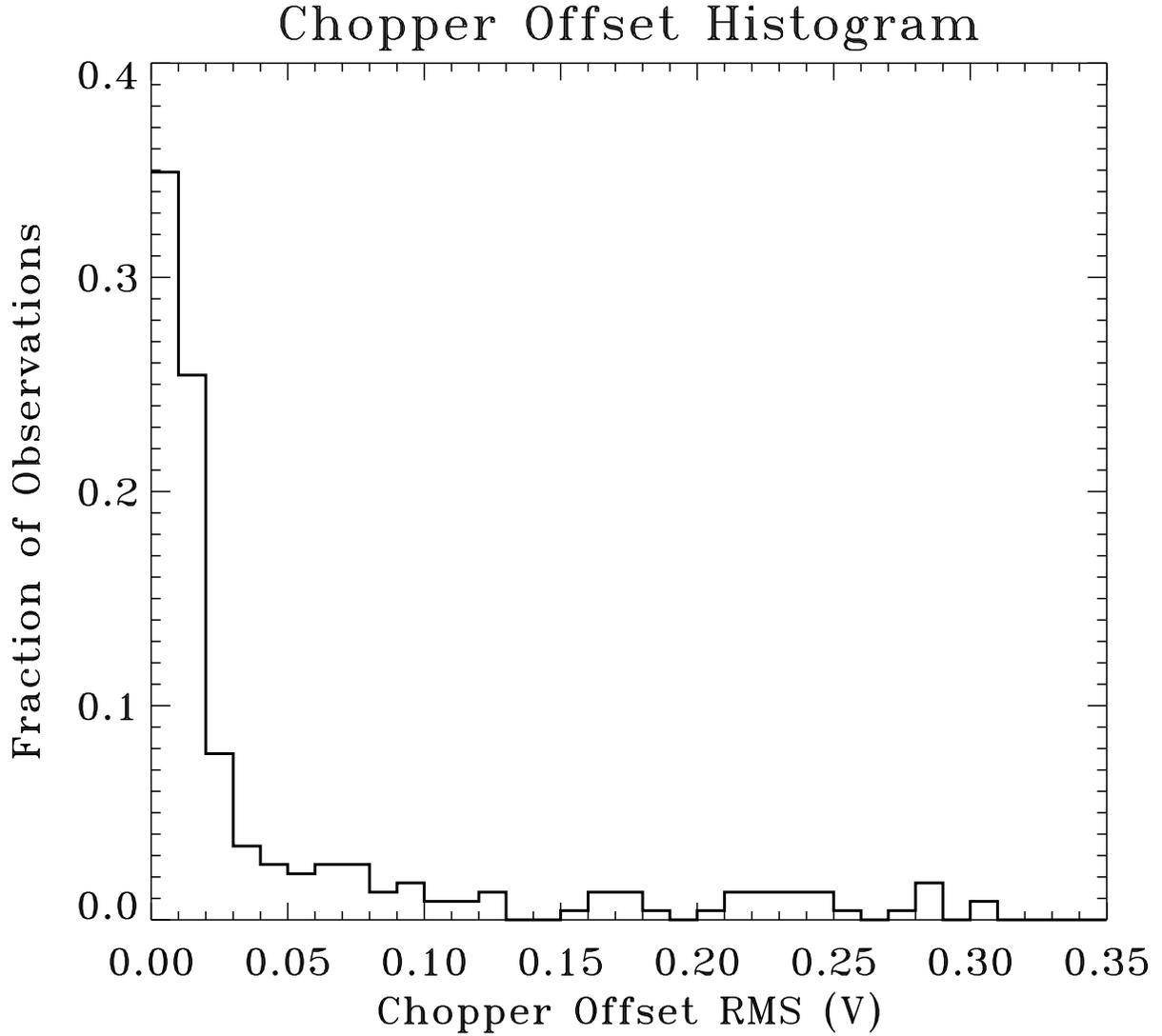}
\caption[Histogram of Chopper Synchronous Offset RMS]{Histogram of 220 GHz chopper
synchronous offset RMS (in volts) during observations of the CMB5 blank field in 2002. 
The approximate calibration from voltage to RJ temperature
at 220 GHz is $0.4~{\rm K\,V^{-1}}$.
The snow cut applied to this data was an RMS less than 0.02 V; this 
conservative cut removed $\sim40\%$ of the data.}
\label{fig:offset_hist}
\end{figure}

Our final data cut is referred to as the ``snow cut."  
As mentioned above in \S\ref{sec:snow}, accumulation of snow
on the telescope mirrors causes a chopper synchronous signal and attenuates
astrophysical signals before they reach the detectors.
The magnitude of signal attenuation depends strongly upon frequency and the
snow cut threshold can be relaxed for an analysis incorporating only the low-frequency data.
Figure \ref{fig:offset_hist} shows a histogram of chopper synchronous offset
RMS during observations of the CMB5 field.  
The snow cut employed for the analysis of the 150 GHz CMB power spectrum 
presented in \citet{kuo02} corresponds
to a chopper offset RMS of 20 mV.  This cut level limits the signal attenuation from snow
to $\lesssim 5\%$ at 150 GHz and removed $\sim40\%$ of the raw data.  
There is a high degree of correlation between the data corrupted by snow accumulation
and the data with increased atmospheric correlation.  This indicates that 
rapid snow accumulation is associated with poor atmospheric observing conditions.

\section{Calibration}\label{sec:calib}

\subsection{Planetary Observations} 

\begin{figure}
\rotatebox{90}{\plotone{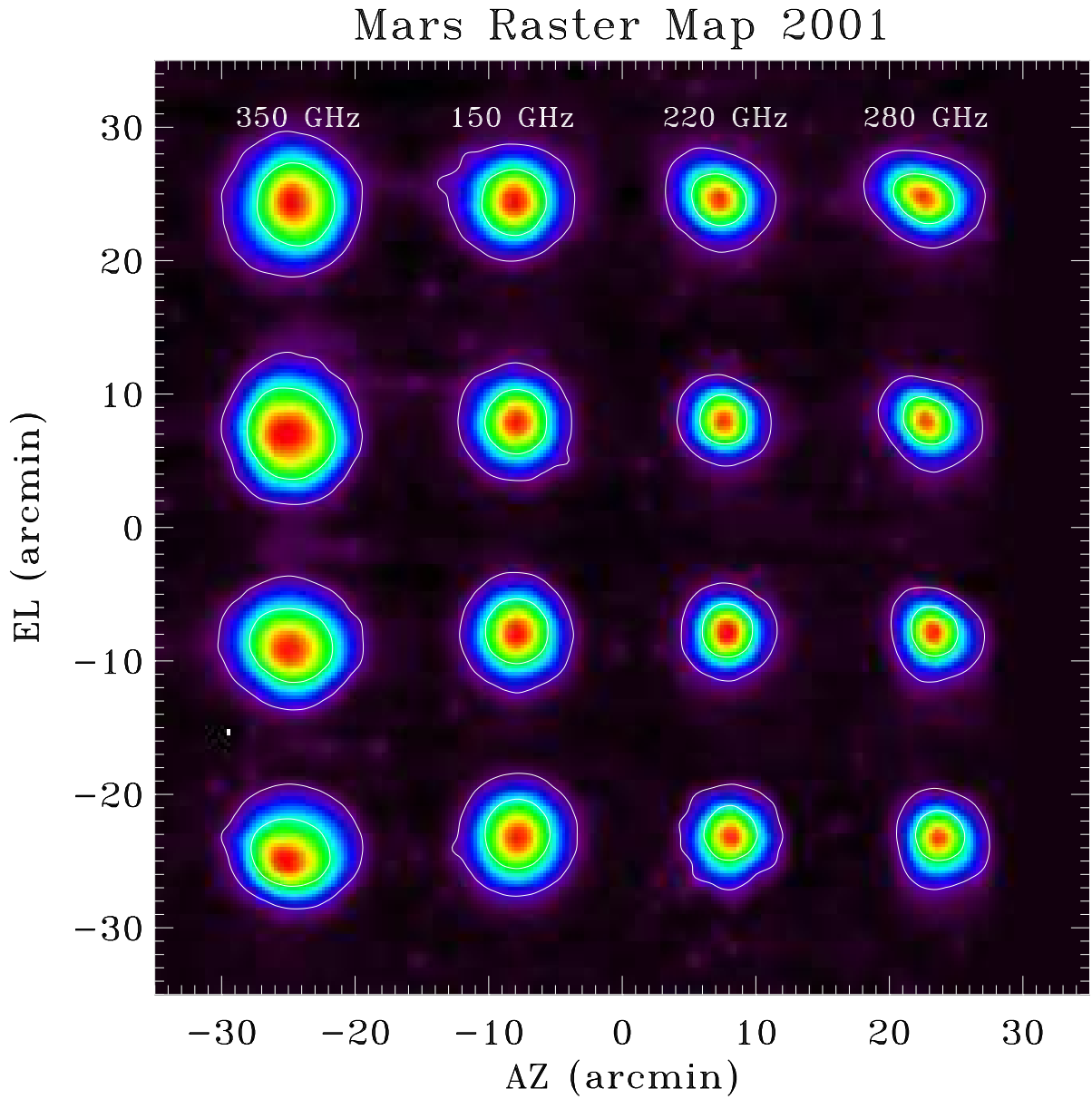}}
\caption[Raster Map of Mars in 2001]{Raster map of Mars at an elevation of 
$26.5^\circ$ taken on 2001 July 16.  The angular diameter of Mars was 
$19^{\prime\prime}$ on this date and the map has been smoothed with a $1^\prime$
FWHM Gaussian.  The white contours correspond to power levels
of 10\% and 50\%.}
\label{fig:mars}
\end{figure}

To convert measured signal voltages to physically
meaningful units, we observe an object of known flux.  Our primary calibration source
for the 2001 season was the planet Mars, which has been well studied at millimeter wavelengths
\citep{griffin86}.  We observed Mars multiple times during the year in an effort to develop a
consistent calibration and check for systematic effects.  For the 2002 season, we used the
planet Venus as our primary calibrator.  Venus was only visible above the horizon
at the end of the 2002 season.
Much less is known about Venus at millimeter
wavelengths and its complicated atmosphere makes the uncertainty of its brightness
temperature quite high ($\sim8\%$ at 150 GHz) \citep{ulich81}.  
Raster maps of Mars and Venus used to calibrate the instrument are shown in 
Figures \ref{fig:mars} and \ref{fig:venus}.

We determine the flux of Mars during an observation using the 
FLUXES\footnote{Available from http://www.starlink.rl.ac.uk} software package
developed for the JCMT telescope on Mauna Kea.  FLUXES provides the
brightness temperature of some of the 
planets on any date across a range of millimeter and sub-millimeter
wavelengths.  FLUXES incorporates a model to correct the Martian brightness
temperature with varying Sun-Mars distance.  For Venus, we use the table of published 
brightness temperatures listed in \citet{weisstein}.  We determine
the position and solid angle, $\Omega_P$, of the planets from the online NASA planetary
ephemeris \citep{nasaephem}. Typical
brightness temperatures for Mars were $\sim208$ K at 150 GHz during our CMB
observations in 2001 with a reported error of $5\%$.  The brightness temperature of Venus is
approximately 294 K at 150 GHz \citep{ulich81} with an error of 8\%.  The planetary
brightness temperatures as a function of frequency, $T_B(\nu)$, are well fit by a straight
line between 100 and 400 GHz.  For each planetary observation, 
we convolve this linear fit with the
frequency response of our detectors, $f(\nu)$, to determine the band-average 
planetary flux density for each channel
\begin{equation}
\bar{S}_P = \frac{\int 2 k T_B(\nu) (\nu/c)^2 \Omega_P f(\nu) d\nu}{\int 
f(\nu)\, d\nu}.
\end{equation}
However, we observe the planet through an attenuating atmosphere and must correct for the
atmospheric opacity to determine the actual planetary flux arriving at the instrument.

\subsection{Atmospheric Opacity}\label{sec:atmopacity}

\begin{figure}
\centering
\plotone{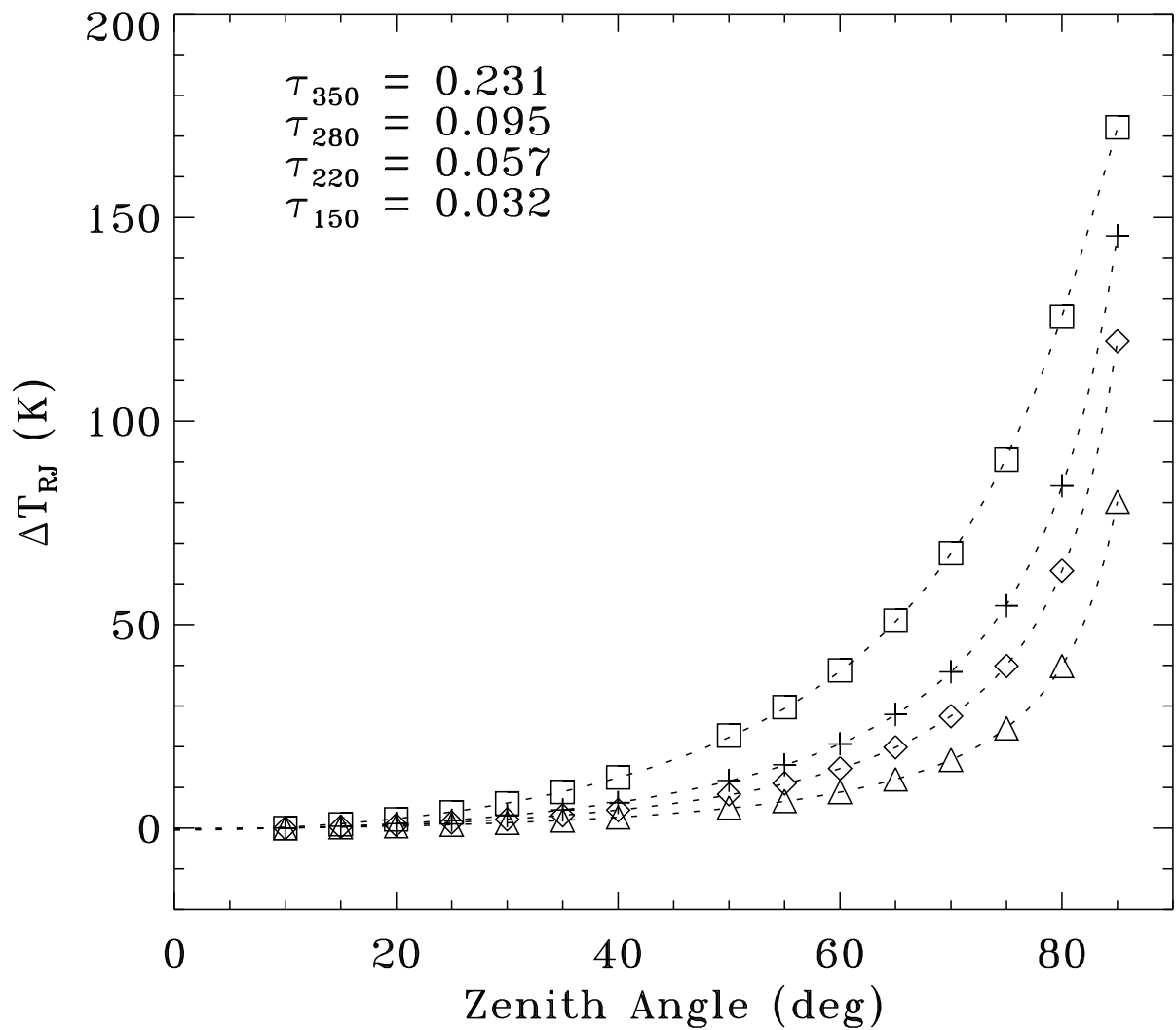}
\caption[Skydip]{Skydip performed on 2001 February 13 when the 350 $\mu$m
zenith opacity was measured to be 1.77.  The figure shows the   
change in sky temperature from the zenith (averaged over the
four channels at each frequency) versus zenith angle ($\zeta$).
The four sets of points are 350 (squares), 280
(pluses), 220 (diamonds) and 150 GHz (triangles) and the dotted lines
are analytic fits to $\Delta T_{RJ}(\zeta) = T_{atm}\left({e^{-\tau} - e^{-\tau/\cos\zeta}}\right)$.  
} 
\label{fig:dip}
\end{figure}

To determine to actual planetary flux incident on our detectors -- as well as 
astrophysical source flux
during normal observations -- we need to characterize the transmission of the atmosphere.  
We perform sky dips to determine the
atmospheric zenith opacity for each channel, $\tau$ (see Figure \ref{fig:dip}).  
Typical measured values of zenith
opacity are 0.033, 0.052, 0.10, and 0.18 at 150, 220, 280, and 350 GHz, respectively.

Because sky dips are a time consuming process, we developed a method which avoids having to
perform sky dips regularly but still permits frequent monitoring of the atmospheric
transmission.  We determined the relationship between the measured
in-band opacities for {\sc Acbar} 
and a 350 $\mu$m tipper experiment located on the adjacent AST/RO building
\citep{radford} which measures the sub-millimeter 
opacity of the South Pole atmosphere approximately
every 15 minutes.  We performed many sky dips during both observing seasons
and correlated the measured in-band zenith
opacities with those measured by the 350 $\mu$m tipper; the relationship is very linear.

\subsection{Detector Responsivity}\label{sec:responsivity}

We also correct for the change in
detector responsivity due to the different optical loadings at low elevation for planets and
high elevation for CMB and cluster observations.  
After each planetary observation, we run the small
calibrator source mounted behind a hole in the tertiary at 5 Hz.  
The chopped source provides a reference signal at planetary elevations
for each channel, $V^{cal}_{P}$, that is used to scale the bolometer responsivity for
future observations.  To correct the change in bolometer responsivity with elevation, we run
the chopped calibrator source again at the elevation of the CMB observations and divide
the responsivity by the measured ratio of the observation calibrator voltage,
$V^{cal}_{obs}$, to the planetary calibrator voltage.  The slow chop frequency
makes the measured calibrator signal insensitive to any changes in detector time
constants with loading.

Because of the gradual accumulation of snow around the hole in the tertiary,
this responsivity transfer is only useful for observations spaced closely in time.  
However, after the snow cut, the measured signal from the galactic source RCW38 
has an RMS scatter of $<4\%$ which is largely due to residual signal attenuation
from the snow.  This also demonstrates that the detector responsivity at high 
elevation is quite stable.  The calibration for each observation is then
\begin{equation}
R = \frac{\bar{S}_P}{\int V_P(\Omega)\,d\Omega}\frac{e^{-\tau_{P}
/cos(\zeta_{P})}}{e^{-\tau_{obs}/cos(\zeta_{obs})}}\frac{V^{cal}_{P}}{V^{cal}_{obs}}
\qquad\text{[Jy V$^{-1}$sr$^{-1}$]},
\end{equation}
where $\int V_P(\Omega)\,d\Omega$ is the integral of the planetary voltage
map over the sky and $\tau$ and $\zeta$ are the atmospheric opacity and zenith 
angle of the observation, respectively.

Observations of the cosmic microwave background are usually calibrated into CMB temperature
units.  Because we measure fluctuations in the CMB (rather than total power)
the conversion between flux density and temperature is given by
\begin{equation}
\Delta S = \frac{dB_{\nu}}{dT}\Omega_B \Delta T,
\end{equation}
where $B_{\nu}$ is the black body spectral energy density and $dB_{\nu}/dT$ is evaluated at
$T_{CMB}$.  The desired calibration from signal voltage to CMB temperature units is given by
\begin{equation}
R^{CMB} = R \left[{\frac{\int\frac{dB_{\nu}}{dT} f(\nu)\, d\nu}{\int f(\nu)
\,d\nu}}\right]^{-1} \qquad\text{[K V$^{-1}$]},
\end{equation}
where we have averaged the conversion from flux density to Kelvin over the
frequency band.  
We estimate the total $1\sigma$ 
uncertainty in the calibration for 2001 and 2002 to be $\pm 10\%$.  

The breakdown of contributions to the 2002 calibration uncertainty at 150 GHz
is $8\%$ from the brightness temperature of Venus, $2\%$ from the residual
RMS in the planetary voltage map integral, 
$2\%$ from calibrator responsivity scaling, and $<1\%$ from the transmission of the atmosphere.
As discussed in the following section, we cross-calibrated the galactic source RCW38
in 2001 and 2002 to check for systematic effects.  We include an additional
calibration uncertainty of $3\%$ from the scatter in cross-calibrated RCW38 flux from
Venus at 150 GHz and $3\%$ systematic uncertainty 
from the consistency of the RCW38 calibration between 2001 and 2002.  Table 
\ref{tbl:systematics} includes a list of potential systematics and their 
contribution to the calibration uncertainty at 150 GHz.

\subsection{Galactic Source Cross-Calibration}\label{sec:crosscal}

During the 2002 observing season, there were no bright planets available until September 2002.
To monitor the calibration during the year, we use the frequent galactic source observations
and cross-calibrate the galactic sources with a planet.  We determine the band-averaged 
integrated flux density of an object, $S_2$, from an object of known flux density, $S_1$, by
\begin{equation}
S_2 = S_1 \frac{\int V_2 d\Omega}{\int V_1 d\Omega} \frac{V_1^{cal}
e^{-\tau_1/\cos\zeta_1}}{V_2^{cal} e^{-\tau_2/\cos\zeta_2}},
\end{equation}
where the ratio $(V_1^{cal} e^{-\tau_1/\cos\zeta_1})/(V_2^{cal} e^{-\tau_2/\cos\zeta_2})$
accounts for the change in atmospheric transmission and bolometer responsivity between the
two observations.

\begin{figure}
\centering
\resizebox{4in}{!}{\plotone{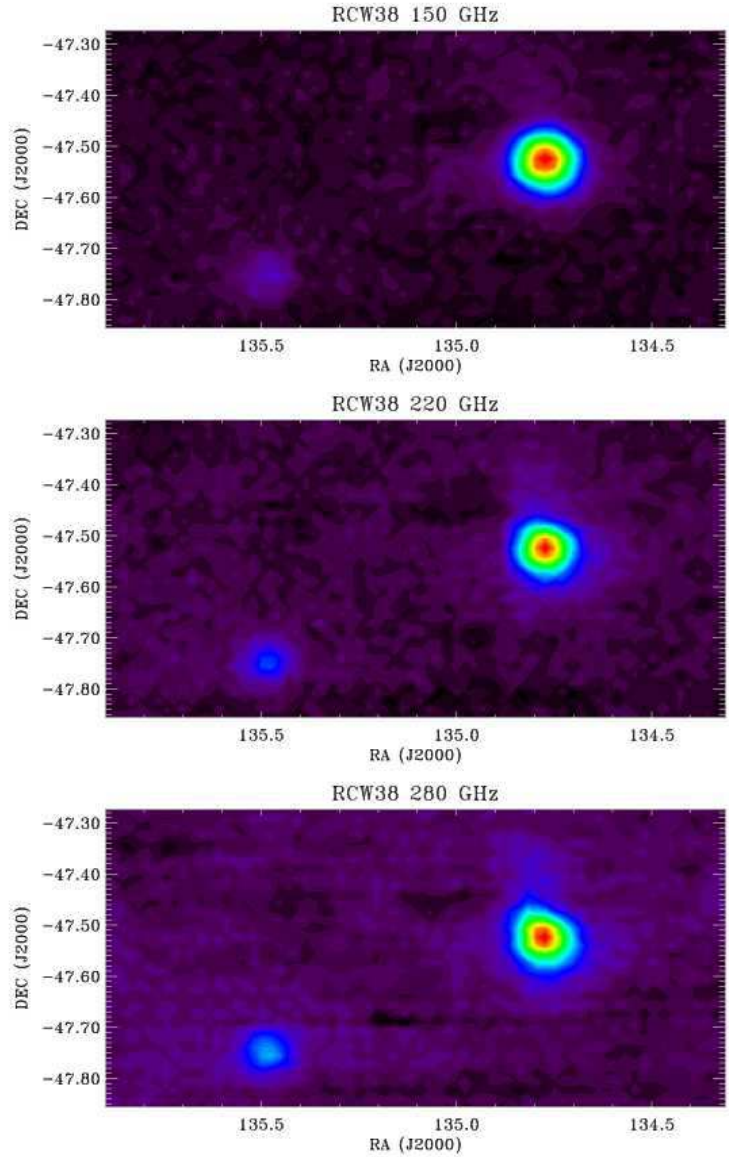}}
\caption[Raster Map of RCW38]{Single-channel raster maps of RCW38 taken on 2001 June 9.
The signal-to-noise on the source is very high in a single raster map.  We use RCW38
to monitor both the telescope pointing and instrument calibration.  The source
in the lower-left corner of the maps appears to be associated with IRAS 09002-4732. } 
\label{fig:rcw38_rast}
\end{figure}

We apply this method to RCW38 in both 2001 and 2002 with Mars and Venus, respectively.
We used RCW38 and planetary observations taken within one day of each other and 
insist that both observations have calibrator runs for scaling the detector
responsivity.  We integrate the flux within $8^\prime$ of the source center; these results
are given in Table \ref{tbl:rcw38boot}.  The agreement between the integrated flux is
good at 150 and 280 GHz but the 220 GHz observations in 2001 suffer from low number
statistics.  A raster map of RCW38 taken on 2001 June 9 is shown in Figure
\ref{fig:rcw38_rast}.  The complex extended structure of RCW38 -- particularly at 220 and
280 GHz -- is apparent even in a single raster image.  

\section{Noise and Sensitivity}\label{sec:noise}

There are multiple sources that contribute to noise in the
{\sc Acbar} instrument \citep{richards94,mather84} but the instrument is fundamentally
limited by the random arrival of background photons.  
Characteristics that determine whether the system will be background limited are: optical
power incident on a detector, bandwidth of observed light, optical efficiency, detector
impedance, detector thermal conductivity, operating temperature, electronics noise, as well
as the atmospheric conditions of the site \citep{griffinholland}.  
The complete noise budget for {\sc Acbar} in 2002 is presented in Table \ref{tbl:noise};
we have included estimates of the noise
contributions from the photon background ($NEP_\gamma$), 
Johnson noise ($NEP_J$), phonon noise ($NEP_G$), and amplifier noise ($NEP_A$).
Refer to \citet{richards94} and \citet{mather84} 
for details on the calculation of noise contributions in bolometric systems.
We have included expected contributions from both the counting and 
photon bunching in $NEP_\gamma$ in our estimated noise table. 
See Appendix \ref{app:noise} for a discussion on the inclusion of
photon bunching.

\begin{figure}
\centering
\plotone{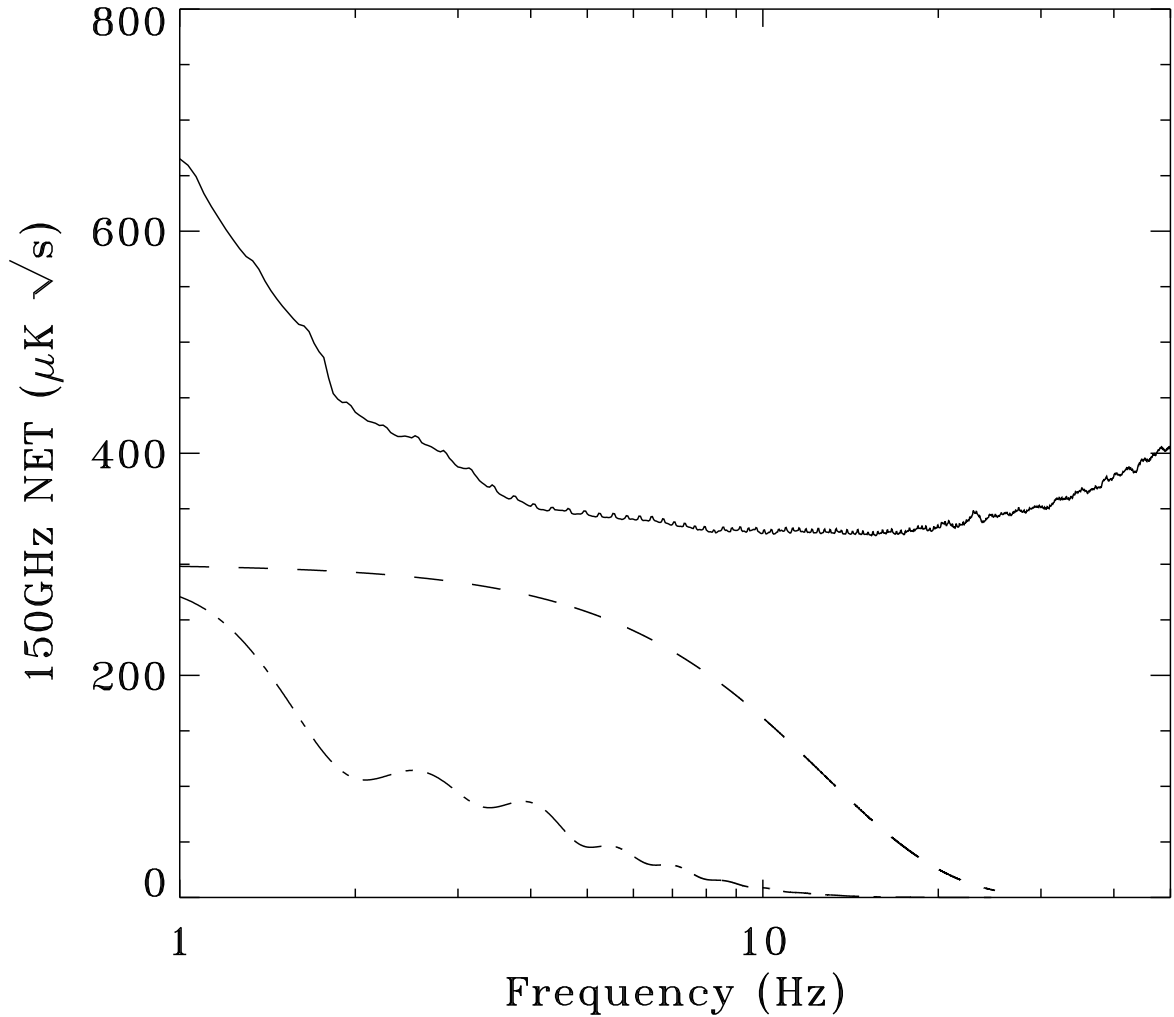}
\caption{Average noise spectrum (in CMB temperature units) at 150 GHz
(solid line) while on the telescope with the chopper running on 2002 May 5.  
Johnson and amplifier noise do not depend upon frequency and result
in a rise in the noise level after dividing by the transfer function.
Also shown are
arbitrarily normalized spectra for a point source (dashed) and a $\Lambda$CDM 
concordance power spectrum (dash-dot) where we have used the chopper velocity for 
a $3^\circ$ chop at 0.3 Hz.  Harmonics of the chopper sweep appear as small
spikes in the noise spectrum.  At this scan velocity, the conversion
from frequency to $\ell$ is approximately $\ell\sim200\nu$ with frequency in Hz.
One can see from the figure that the peak sensitivity 
for this chopping speed overlaps with the damping tail of the CMB power
spectrum and the $1/f$ rises in for the largest angular scales where {\sc Acbar} is
dominated by cosmic variance.}
\label{fig:net150}
\end{figure}

The average total $NET_{RJ}$ between 10 and 20 Hz (above the $1/f$ knee
from the electronics and atmosphere but well within the signal band) for 2002 were
200, 250, and 280 $\mu{\rm K}\sqrt{\rm s}$ at 150, 220, and 280 GHz, respectively.  
For comparison, the $NET_{RJ}$ on a very good day in 2001 
were 340, 380, 270, and 950 $\mu{\rm K}\sqrt{\rm s}$ at
150, 220, 280, and 350 GHz, respectively.  The substantial improvement at the 
lower frequencies resulted from a filter upgrade in December 2001.  When combined
with the addition of a second row of 150 GHz channels (replacing the 350 GHz
detectors), the upgrade resulted in
a nearly six-fold improvement in 150 GHz mapping speed.
A plot of the average $NET_{CMB}$ versus frequency for the 150 GHz channels in
2002 is shown in Figure \ref{fig:net150}. 

\section{Conclusions}\label{sec:conc}

We have described the design and performance of the Arcminute Cosmology
Bolometer Array Receiver ({\sc Acbar}).  We have taken advantage 
of improvements in bolometric detector technology and the superb observing
conditions afforded by the South Pole to make very sensitive, high resolution
maps of the CMB at multiple millimeter wavelengths.  {\sc Acbar} was installed on
the Viper telescope in January 2001 and has mapped the microwave sky through
November 2002.  The instrument is background photon limited at all frequencies, 
with an average $NET_{CMB}$ per detector of $350~\mu{\rm K}\sqrt{\rm s}$ at 150 GHz,
and the measured noise level is more consistent with the noise model if
we include the photon bunching contribution.

The CMB power spectrum measurement from a subset of {\sc Acbar}'s 2001 and 2002 data is 
presented in \citet{kuo02} and the constraints on cosmological parameters 
derived from
this power spectrum are given in \citet{goldstein02}.  
We have searched within
these maps for SZ clusters of galaxies \citep{runyan03b} to constrain models
of structure formation.  As part of a multi-wavelength investigation of nearby
cluster physics, we have also surveyed a fluxed-limited sample of
clusters at $z<0.1$ with sufficient sensitivity at multiple frequencies
to separate the primary CMB and thermal SZ emission \citep{gomez02}.

We acknowledge assistance in the fabrication of {\sc Acbar} by the 
capable staff of the 
U.C. Berkeley, U.C. Santa Barbara, and Caltech machine shops
and the U.C. Berkeley electronics shop. 
The support of Center for Astrophysics Research in Antarctica (CARA) 
polar operations has been essential in the installation and operation of 
the telescope.  We would like to thank John Carlstrom and Steve Meyer
for their early and continued support of the project.
We gratefully acknowledge Simon Radford for 
providing the 350 $\mu$m tipper data.  We thank Charlie Kaminski
and Michael Whitehead for their assistance with winter observations.
This work has been primarily supported by NSF Office of Polar
Programs grants OPP-8920223 and OPP-0091840.  Marcus Runyan
acknowledges many useful discussions with B.~P. Crill, S. Golwala,
W.~C. Jones, P.~D. Mauskopf, and B.~J. Philhour, help with lab
measurements by Dan Marrone, and the support 
of a NASA Graduate Student Researchers Program fellowship.  
Chao-Lin Kuo acknowledges support 
from a Dr. and Mrs. CY Soong fellowship.  We also thank the referee
for useful suggestions on the manuscript.

\appendix

\section{Bandwidth and Optical Efficiency}\label{app:band}

In this appendix we discuss the subjects of system bandwidth and optical efficiency. 
Traditionally, the optical bandwidth of a system is defined as the full-width at
half the maximum transmission of a measured frequency spectrum.  Although this number
is quantitative in the sense that it can be measured precisely from any given spectrum, it
does depend intrinsically upon the resolution of the spectrum from which it is
calculated.  For example, fringes in a high-resolution spectrum will tend to
push the ``half-maximum'' point further up the band compared to a smooth 
low-resolution spectrum from the same optical system.

Consider an optical system with measured spectral response, $f(\nu)$.  This 
spectrum is normalized by measuring the multiplicative optical
efficiency coefficient, $\eta$,
such that the optical power, $Q$, measured from a 
beam-filling source of spectral intensity, $I(\nu)$, is given by
\begin{equation}
Q = \eta \int f(\nu)I(\nu)A(\nu)\Omega(\nu) d\nu,
\end{equation}
where $A(\nu)\Omega(\nu)$ is the frequency-dependent optical throughput of
the system; $A\Omega = \lambda^2$ in the case of a single-moded system.  We
define the efficiency normalized spectral response as $\tilde{f}(\nu) = \eta f(\nu)$.
In the case of a RJ emitting source at temperature $T$, the spectral intensity 
is given by $I(\nu) = 2kT\nu^2/c^2$ and the optical power absorbed by the detector 
in a single-moded system is $Q = 2kT\int \tilde{f}(\nu) d\nu$.

We would like to identify the quantities bandwidth, $\Delta\nu$, and average
optical
efficiency, $\bar{\eta}$, such that $Q = 2NkT\bar{\eta}\Delta\nu$, where $N$
is the number of optical modes supported by the system.  Limiting ourselves
to the case of a single optical mode ($N=1$) this leads to the 
relation $\int \tilde{f}(\nu) = \bar{\eta}\Delta\nu$.  The key point is that
the average optical efficiency and effective bandwidth are not separable 
quantities.  As long as the bandwidth is clearly defined, then the average
optical efficiency is given by $\bar{\eta} = \Delta\nu^{-1}\int \tilde{f}(\nu) d\nu$.
Although the width of the band at half maximum will give a ball park measurement
of the bandwidth, we would like to use a bandwidth estimator that is robust to changes
in the resolution of the spectrum.  

The derivative of a transmission spectrum with frequency, 
$\tilde{f}^\prime = d\tilde{f}/d\nu$, 
emphasizes those regions of the spectrum where
the band is changing significantly -- it has a large positive value when the
band is rising and a large negative value when the band is falling.  Using the 
derivative of the band as a weight, we can determine the edges of the band.  
High-resolution spectra will result in fringing in the derivative but these
fringes will average out.

Consider an arbitrarily normalized transmission spectrum, $f(\nu)$, with band
center $\nu_0 = \frac{\int \nu f(\nu) d\nu}{\int f(\nu) d\nu}$ and frequency
derivative, $f^\prime (\nu)$.  We identify the lower and upper band edges as
\begin{equation}
\nu_L = \frac{\int_0^{\nu_0} \nu f^\prime (\nu) d\nu}{\int_0^{\nu_0} f^\prime (\nu) d\nu}, \quad
\nu_U = \frac{\int_{\nu_0}^\infty \nu f^\prime (\nu) d\nu}{\int_{\nu_0}^\infty f^\prime (\nu) d\nu}.
\end{equation}
The bandwidth is then defined as $\Delta\nu = \nu_U - \nu_L$.  This bandwidth
estimator depends only weakly upon the exact value of $\nu_0$ used in the integrals
for reasonably contiguous bands ({\it eg.} bands with only have one main ``lobe'').

To see that this definition of bandwidth is sensible, consider the two cases of 
a top-hat and Gaussian band.  In the case of the top-hat band, the derivatives
become delta functions at the band edges and the result is straight forward.  For the
Gaussian case, the band does not have well defined ``edges'' and the derivative 
will have maxima and minima at the inflection points of the Gaussian.  
Even for a Gaussian band with much non-localized transmission, calculating
the bandwidth with the equations above results in a bandwidth a few percent 
larger than the FWHM of the band.

\begin{figure}
\centering
\resizebox{4in}{!}{\plotone{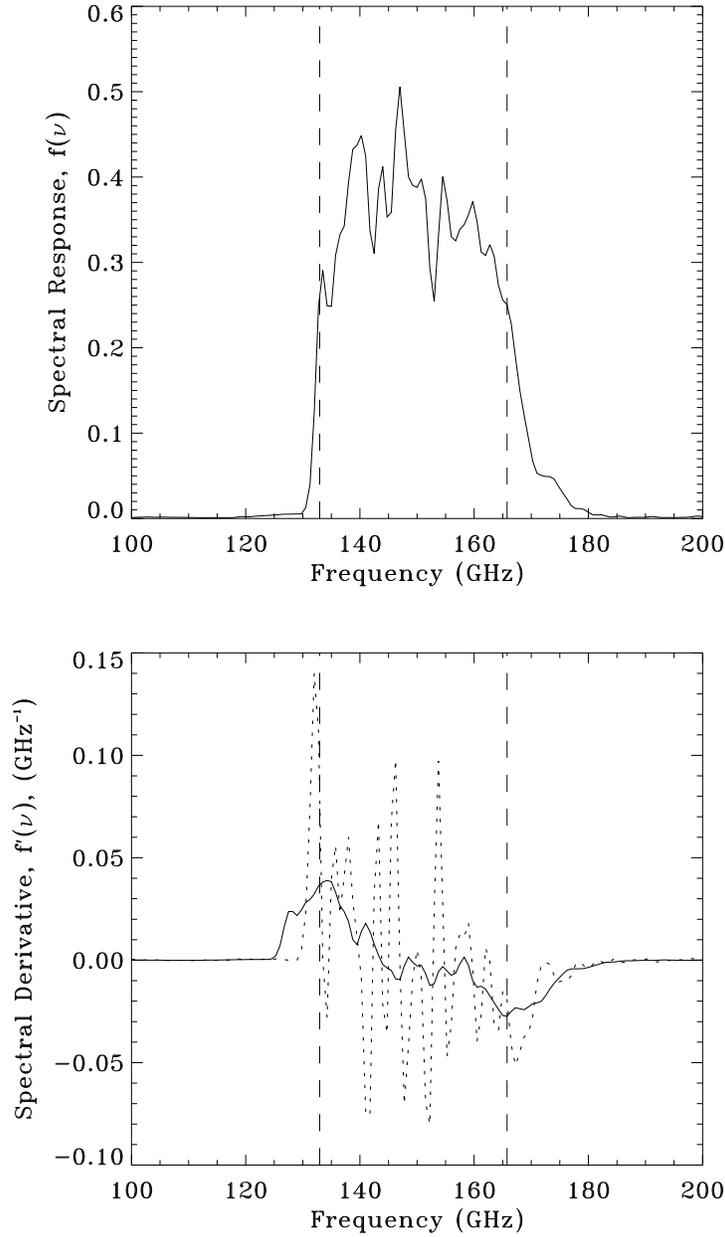}}
\caption{Top panel shows the measured spectral response, $\tilde{f}(\nu)$, 
from one of the
150 GHz {\sc Acbar} channels.  The spectrum has been normalized by the optical
efficiency such that the optical power absorbed by the detector 
from a RJ source is given by $Q=2kT\int \tilde{f}(\nu) d\nu$.  The bottom
panel shows the raw frequency derivative of the spectral response, $f^\prime(\nu) =
df/d\nu$, (dotted) as well as the smoothed derivative (solid) to emphasize
those frequencies where the weighting is largest.  The dashed lines in both figures
are the band edges as determined by the equations presented in Appendix \ref{app:band}. }
\label{fig:bandwidth}
\end{figure}

Real-world bands are not so nicely behaved as the two examples above.  In 
Figure \ref{fig:bandwidth} we show the average of the {\sc Acbar} 150 GHz bands as
well as its derivative.  We show both the raw derivative as well as a smoothed
derivative to show where the derivative will weight the band edges most heavily.
Because the frequency, $\nu$, is a smoothly varying function, the rapid variations
in the derivative are averaged out in the integral.  
This method was used to calculate the bandwidths and average optical efficiencies
presented in this paper and should yield normalization and resolution independent measures
of bandwidth
for well behaved spectral bands ({\it eg.} bands that do not have bulk features
such as a large gouge in the middle of the band).

\section{Photon Noise}\label{app:noise}

Photon noise is the result of temporal fluctuations in the rate of photon 
arrival at the detectors. A series of pioneering papers by 
\citet{hanburybrown57,hanburybrown58a,hanburybrown58b}
showed that in addition to the Poisson counting noise,
there is a bunched component due to the Bose statistics
describing the photons.
The same authors also showed that if the detection bandwidth is much smaller 
than the optical bandwidth ($30$ Hz {\it vs.} $30$ GHz for {\sc Acbar}), both 
noise components have a white power spectrum \citep{hanburybrown57}. 

Following the derivation given in Appendix A of \citet{hanburybrown58a},
\citet{lamarre} finds an expression for the noise equivalent power ($NEP$) 
of the fluctuations in optical power; 
\begin{equation}
NEP^2=\left[2\int h\nu Q_\nu d\nu +
(1+\!P^2)\int {\mathcal D}_\nu Q_\nu^2 d\nu\right]\qquad
\left[\rm W^2 Hz^{-1}\right],
\label{photonpsd}
\end{equation}
where $P$ is the degree of polarization, $Q_\nu$ is the specific flux
(${\rm W Hz^{-1}}$), and the partial coherence factor ${\mathcal D}_\nu$ is 
\begin{equation}
{\mathcal D}_\nu\equiv \frac{1}{Q_\nu^2}\int\!
\frac{d^2\xi d^2\xi'd^2xd^2x'}{R^4}
\sqrt{I_\nu({\bf x},{\boldsymbol \xi})I_\nu({\bf x'},{\boldsymbol \xi})
I_\nu({\bf x},{\boldsymbol \xi'})I_\nu({\bf x'},{\boldsymbol \xi'})}\cos\left\{
\frac{2\pi\nu}{cR}\left[({\bf x}-{\bf x'})
\cdot({\boldsymbol \xi}-{\boldsymbol \xi'})\right]\right\}\;.\label{deltanu}
\end{equation}
${\boldsymbol \xi},{\boldsymbol \xi'}$ are points
on the source and ${\bf x},{\bf x'}$ are points on the detector.
$R$ is the distance to the source in the far field. 
$I_\nu({\bf x},{\boldsymbol \xi})$ is the optical intensity field. 
For a given source/detector geometry, ${\mathcal D}_\nu$ represents the degree
of second-order coherence (intensity coherence). It is very similar 
to van Cittert-Zernike's degree of coherence \citep{born_wolf},
the only difference being that the latter is defined for the first order 
coherence (amplitude coherence).

There has been considerable confusion as to the correct form of the
photon noise. It was argued that the second term in equation (\ref{photonpsd}) 
(the bunching term) should be further suppressed due to the loss 
of temporal coherence [See \citet{richards94} for review]. We show that
this is not the case, since equation (\ref{photonpsd}) directly leads 
to the correct expression for fluctuations in photon counts. 

If $Q_\nu$ does not vary significantly within the IR bandwidth $\Delta\nu$,
the average number of photons detected in time $\Delta t$ is 
\begin{equation}
\langle N\rangle=\frac{Q_\nu\Delta \nu \Delta t}{h\nu}.
\end{equation} 
The photon noise is white, and the variance of 
$N$ can be easily calculated from the $NEP$,
\begin{equation}
\langle (\Delta N)^2\rangle=\frac{\Delta t^2({NEP}^2\Delta B)}{h^2\nu^2}=
2(\Delta B\;\Delta t)\left[\langle N\rangle+\frac{1+P^2}{2}
\frac{{\mathcal D}_\nu\langle N\rangle^2}
{\Delta t\Delta\nu}\right]\equiv\langle N\rangle+\frac{\langle N\rangle^2}
{g}\;.\label{nn1}
\end{equation}
 Here $\Delta B$ is the signal bandwidth, $\Delta B=1/2\Delta t$, and 
\begin{equation}
g\equiv \frac{2}{1+P^2}\frac{\Delta\nu\Delta t}{{\mathcal D}_\nu}\;,\label{g}
\end{equation}
For a polarized ($P=1$) point source (${\mathcal D}_\nu=1$, see below), 
$g$ reduces to $\Delta\nu\Delta t$. In that limit, 
equation (\ref{nn1}) is in agreement with results derived by Jakeman and Pike
\citep{jakeman68,loudon} for photon counting statistics. On the other hand, 
the mean square of the {\it instantaneous}
photon occupation number $n$ for blackbody radiation is
\begin{equation}
\langle (\Delta n)^2\rangle=\langle n\rangle+\langle n\rangle^2\;.\label{nn2}
\end{equation}
Compared with this standard result, there is an extra factor of $g$ 
in equation (\ref{nn1}). This extra factor should not be interpreted 
as an averaging effect and should not be applied to the second term of 
equation (\ref{photonpsd}). 
Since the factor $g$ does {\it not} reduce 
the relative contribution from the bunching term even in the limit that 
$(\Delta\nu\Delta t)\gg 1$. The ratio of the Bose 
term to the Poisson term is fixed by frequency and independent of integration 
time or bandwidth:
\begin{equation}
\frac{{\rm Bose}}{{\rm Poisson}}=
\frac{\langle N\rangle}{g}=\frac{Q_\nu{\mathcal D}_\nu}{h\nu}
\frac{1+P^2}{2}\;.
\end{equation}
However in multi-moded systems $\mathcal D_\nu<1$, and the bunching term is 
suppressed. 

\citet{lamarre} shows that for a point source, ${\mathcal D}_\nu=1$; in the
other limit, if the source fills the beam and the throughput $U$ of 
the detector is much greater than the coherence etendue $c^2/\nu^2$ 
(diffraction-limited throughput), equation (\ref{deltanu}) reduces to
\begin{equation}
{\mathcal D}_\nu=\frac{c^2}{U\nu^2}.\label{largeu}
\end{equation}
Note that this equation is only applicable when ${\mathcal D}_\nu\ll 1$.
The physical interpretations of these two limits are the following.
In the case of a point source, the photons received across the
detector are in perfect spatial coherence and exhibit maximal 
bunching. On the other hand, if the light source is 
extended, not only the first order coherence (van Cittert-Zernike 
integral), but the second order coherence (${\mathcal D}_\nu$ defined above) 
decreases due to cancelation. The decrease in first/second order 
coherence was used by Michelson/Hanbury Brown \& Twiss to determine 
the apparent size of stars \citep{jakeman70}.

For a diffraction-limited experiment like {\sc Acbar},
the exact value of ${\mathcal D}$ requires explicit evaluation
of equation (\ref{deltanu}) for all sources of optical loading 
(filters, sky, telescope, etc.). 
We have not performed this imposing calculation, but expect that 
${\mathcal D}_\nu$ should be of order unity and 
lie in the interval $0<{\mathcal D}_\nu<1$. 
This qualitatively explains why $NEP_{\rm w/o\;bunching}
<NEP_{\rm achieved}<NEP_{\rm w/\;bunching}$ (Table \ref{tbl:noise}).

\bibliographystyle{apj}
\bibliography{merged}

\clearpage


\clearpage


\begin{table*}
\begin{center}
\caption{Feed Dimensions and Measured and Expected Beam Widths}
\label{tbl:beamparams}
\begin{tabular}{|c|c|c|c|c|c|c|c|}
\hline
Frequency & $a$ & $L$ & $d$ & $w$ & $p$ & Gaussian & Model \\
(GHz) & (mm) & (mm) & (mm) & (mm) & (mm) & Fit FWHM & FWHM \\
\hline\hline
150 & 6.1 & 35.0 & 0.80 & 0.72 & 1.08 & $12.52^\circ\pm 0.01^\circ$ & $12.57^\circ$ \\
220 & 6.1 & 45.0 & 0.54 & 0.48 & 0.72 & $8.65^\circ\pm 0.03^\circ$ & $8.66^\circ$ \\
280 & 6.2 & 53.0 & 0.44 & 0.40 & 0.60 & $6.79^\circ\pm 0.08^\circ$ & $6.81^\circ$ \\
350 & 5.9 & 22.1 & 0.34 & 0.36 & 0.54 & $5.14^\circ\pm 0.02^\circ$ & $5.27^\circ$ \\
\hline
\end{tabular}
\end{center}
\tablecomments{Physical dimensions
of feeds: $a$ is the radius of feed aperture, $L$ is the length of the conical section,
$d$ is the depth of the corrugations in the conical section ($\lambda/4$), 
$w$ is the width of the groves ($2\lambda/3$), and $p$ is the groove pitch ($\lambda/3$).
The width and pitch of the grooves for the 350 GHz channels were made slightly larger
than the nominal values for ease of manufacture.
The 350 GHz feed is lensed and does not follow the length progression with frequency.
Also listed are the average measured Gaussian FWHM and model
Gaussian FWHM for all four {\sc Acbar} frequency feeds.  The error bars are the statistical
scatter of the four measured beam FWHMs within a frequency.}
\end{table*}

\begin{table*}
\begin{center}
\caption{2002 Filter Arrangement}
\label{tbl:filters}
\begin{tabular}{|c|c|c|c|c|}
\hline
Filter & Temp (K) & 150 GHz & 220 GHz & 280 GHz \\
\hline\hline
Blocker \#1 & 240 mK & 255 GHz & 360 GHz & 420 GHz \\
Edge & 240 mK & 169 GHz & 235 GHz & 300 GHz \\
alkali-halide & 4 K & 1.6 THz & 1.6 THz & 1.6 THz \\
Pyrex & 4 K & 1.2 THz (0.019$^{\prime\prime}$) & 1.2 THz (0.02$^{\prime\prime}$) & 1.2 THz (0.021$^{\prime\prime}$) \\
Blocker \#2 & 4 K & 234 GHz & 360 GHz & 420 GHz \\
Waveguide & 4 K & 131 GHz (0.053$^{\prime\prime}$) & 204 GHz (0.03$^{\prime\prime}$) & 252 GHz (0.0275$^{\prime\prime}$) \\
Blocker \#3 & 77 K & 420 GHz & 420 GHz & 420 GHz \\
\hline
\end{tabular}
\end{center}
\tablecomments{Filter elements of {\sc Acbar} in the 2002 configuration.  Frequencies denote
approximate 50\% transmission points of blocker, edge, and waveguide filters, but corresponds
to roughly 10\% points of Pyrex and alkali-halide (AH).  
Numbers in parentheses are thicknesses for Pyrex
and diameter for the waveguide.  The 77 K blocker is a single, large, 
metal-mesh filter mounted on
the nitrogen can that all of the feeds look through.  This upgraded filter arrangement
differs from the 2001 configuration which did not have Blockers \#2 or \#3 or the Pyrex, but
instead had black-poly disks at 4 K. In 2001 the AH filter was located in front of
the feed structure at 77 K and contributed significantly to the internal loading of the
system; it was replaced with smaller AH disks within the feeds at 4 K for 2002. }
\end{table*}

\begin{table*}
\begin{center}
\caption{Measured Optical, Electrical, and Thermal Bolometer Properties}
\label{tbl:boloparams}
\begin{tabular}{|c|c|c|c|c|c|c|c|}
\hline
Freq & $R_0$ &  $G_0 $ & $\beta$ & $\nu_0$ & $\Delta\nu$ & $\bar\eta$ & $\tau$ \\
(GHz)& $(\Omega)$ & (pW $K^{-1}$) & $~$ &	(GHz) &	(GHz) & (\%) & (ms) \\
\hline\hline
150 & 150 & 376 & 1.11&	151& 31 & 40 & 3.6 \\
220 & 143 & 403 & 1.14& 222& 31 & 34 & 5.8 \\
280 & 138 & 630 & 1.15& 283& 48 & 33 & 1.8 \\
350 & 165 & 651 & 1.06& 352& 24 & 28 & 0.9 \\
\hline
\end{tabular}
\end{center}
\tablecomments{Average {\sc Acbar} bolometer electrical and
optical properties for 150, 220, and 280 GHz channels in 2002.  There
were no 350 GHz detectors installed 
on the focal plane in 2002, so we include the values from 2001 for 350 GHz in
the table.
The detector impedance follows $R(T)=R_0 e^{\sqrt{\Delta/T}}$, where $\Delta$
depends on the doping of the NTD thermistors and is 41.8 K for all
of the bolometers.  The functional form of the thermal conductivity is 
$G(T) = G_0 (T/T_0)^\beta$ where $G_0$ are normalized to $T_0 = 300$ mK.  
$\nu_0$ is the measured
band center, $\Delta\nu$ is the bandwidth, and $\bar{\eta}$ is the
effective optical efficiency across the band (see Appendix \ref{app:band}
for definitions).  $\tau$ is the effective
optical time constants under operating conditions.}
\end{table*}

\begin{table*}
\begin{center}
\caption{Beam Sizes}
\label{tbl:beams}
\begin{tabular}{|c|c|c|c|c|}
\hline
Year & 150 GHz  & 220 GHz  & 280 GHz & 350 GHz  \\
\hline\hline
2001 & $4.80^\prime\pm0.17^\prime$ & $4.00^\prime\pm0.21^\prime$ & 
   $3.97^\prime\pm0.30^\prime$ & $5.66^\prime\pm0.44^\prime$  \\
2002 & $4.69^\prime\pm0.15^\prime$ & $4.25^\prime\pm0.36^\prime$ & $3.94^\prime\pm0.25^\prime$ & -- \\
\hline
\end{tabular}
\end{center}
\tablecomments{Average measured FWHM beam sizes measured from Mars on 2001 July 17 and 
Venus on 2002 September 22. The FWHM quoted in the table is calculated from the measured solid angles
assuming a Gaussian shape with $\theta_{FWHM} = \sqrt{(4\ln 2)\Omega/\pi}$.  Also listed
is the $1\sigma$ 
dispersion in beam sizes for the feeds at each frequency (eight at 150 GHz and 
four at 220 and 280 GHz).  The uncertainty in the individual measured beam 
solid angles is $\sim 2\%$ which is smaller than the feed-to-feed dispersion
in the beam sizes.  There were no 350 GHz feeds in 2002.}
\end{table*} 


\begin{table*}
\begin{center}
\caption{Noise Budget for 2002}
\label{tbl:noise}
\begin{tabular}{|r|c|c|c|}
\hline
Frequency (GHz) & 150 & 220 & 280 \\
\hline\hline
$\Delta\nu$ (GHz) & 31 & 31 & 48 \\
$\bar{\eta}$ (\%) & 40 & 34 & 33 \\
$FWHM$ ($\prime$) & 4.7 & 4.2 & 3.9 \\
$Q_{\rm total}$ (pW) & 13.3 & 10.7 & 29.2 \\
$T_{RJ}$ (K) & 39 & 37 & 65 \\
$R$ (M$\Omega$) & 6.9 & 7.4 & 7.1 \\
$T_{\rm bolo}$ (mK) & 360 & 350 & 360 \\
$G(T)$ ${\rm pW~K^{-1}}$) &  460 & 480 & 760 \\
$S$ ($10^8~{\rm V~W^{-1}}$) & -2.4 & -2.5 & -1.8 \\
$NEP_A~(10^{-17}~{\rm W\,Hz^{-1/2}})$ & 1.3 & 1.2 & 1.6 \\
$NEP_J~(10^{-17}~{\rm W\,Hz^{-1/2}})$ & 2.5 & 2.4 & 3.4 \\
$NEP_G~(10^{-17}~{\rm W\,Hz^{-1/2}})$ & 4.5 & 4.5 & 5.7 \\
$NEP_{\gamma{\rm counting}}~(10^{-17}~{\rm W\,Hz^{-1/2}})$ & 5.2 & 5.7 & 10.3 \\
$NEP_{\gamma{\rm bunching}}~(10^{-17}~{\rm W\,Hz^{-1/2}})$ & 7.1 & 5.8 & 12.1 \\
$NEP_{\rm total~w/o~bunching}~(10^{-17}~{\rm W\,Hz^{-1/2}})$ & 7.4 & 7.7 & 12.4 \\
$NEP_{\rm total~w/~bunching}~(10^{-17}~{\rm W\,Hz^{-1/2}})$ & 10.3 & 9.7 & 17.3 \\
\hline
$NEP_{\rm achieved}~(10^{-17}~{\rm W\,Hz^{-1/2}})$ & 9.4 & 9.5 & 14.8 \\
$NET_{CMB}$ ($\mu K \sqrt{\rm s}$) & 350 & 770 & 1550 \\
$NET_{RJ}$ ($\mu K \sqrt{\rm s}$) & 200 & 250 & 280 \\
$NEFD$ (mJy $\sqrt{\rm s}$) & 290 & 640 & 990 \\
\hline
\end{tabular}
\end{center}
\tablecomments{Average bolometer parameters and noise budget for all three 
frequencies during observations of the CMB5 field (elevation $55^\circ$) 
in 2002. Listed are the average
bandwidth, $\Delta\nu$, band-average optical efficiency, $\bar{\eta}$, fit Gaussian beam
FWHM, total optical power, $Q_{\rm total}$, RJ temperature loading, $T_{RJ}$, 
detector impedance, $R$, bolometer temperature, $T_{\rm bolo}$, detector thermal
conductivity, $G(T)$, and DC responsivity, $S$.  The electronics 
noise contribution is $\sim3~{\rm nV\,Hz^{-1/2}})$ 
and is scaled to $NEP_A$ by dividing by the responsivity, $S$.
The photon counting term (or Poisson term) is equal to $NEP_{\gamma\rm{counting}}^2
= \int{2h\nu Q_\nu d\nu}$ and
the photon bunching term (or Bose term) is equal to 
$NEP_{\gamma\rm{bunching}}^2 = \int{Q^2_\nu d\nu}$.  $Q_\nu$ is the
spectral power absorbed by the bolometer and is assumed to be
$Q_\nu = 2 \tilde{f}_\nu kT_{RJ}\nu^2 A(\nu)\Omega(\nu)/c^2$, 
where $\tilde{f}_\nu$ is the 
optical efficiency normalized spectral response, and the total
loading on the detector is $Q_{\rm total}=\int Q_\nu d\nu$.
We have calculated the expected noise $NEP$s with and without
the bunching term (assuming $\mathcal D_\nu=1$, see 
Appendix~\ref{app:noise} for a discussion).
The total $NEP$ is the quadrature sum of all noise components listed.  
The achieved $NEP$s at the bolometer are derived from calibrated telescope noise 
data between 10 and 20 Hz
with the chopper running; these frequencies are above the $1/f$ knee of
the electronics and atmosphere but well within the signal band.  
}
\end{table*}

\begin{table*}
\begin{center}
\caption{CMB Field Central Point Sources}
\label{tbl:quasars}
\begin{tabular}{|c|c|c|c|}
\hline
 FIELD & name & $\alpha$ (J2000) & $\delta$ (J2000) \\
\hline\hline
CMB2 & PMN J0455-4616 & $73.962^\circ$ & $-46.266^\circ$ \\
CMB5 & PMN J0253-5441 & $43.372^\circ$ & $-54.698^\circ$ \\
CMB6 & PMN J0210-5101 & $32.692^\circ$ & $-51.017^\circ$ \\
CMB7 & PMN J2235-4835 & $338.805^\circ$ & $-48.600^\circ$ \\
\hline
\end{tabular}
\end{center}
\tablecomments{Central pointing reference sources 
for the CMB observations.  The source names and positions are from the 
Parkes-MIT-NRAO (PMN) Radio Survey as reported by the NASA/IPAC Extragalactic 
Database. }
\end{table*}

\begin{table*}
\begin{center}
\caption{Bootstrapped Flux of RCW38}
\label{tbl:rcw38boot}
\begin{tabular}{|c|c|c|}
\hline
 Freq (GHz) & $S_{\text{RCW38}}^{2001}$ (Jy) & $S_{\text{RCW38}}^{2002}$ (Jy) \\
\hline\hline
150 & $146\pm 15$ & $144\pm 14$  \\
220 & $229\pm 30$ & $260\pm 26$  \\
280 & $318\pm 32$ & $323\pm 32$  \\
\hline
\end{tabular}
\end{center}
\tablecomments{Cross-calibrated integral flux densities for RCW38 within 
$8^\prime$ radius for 2001 and 2002.  The 10\% error bars on $S_{\text{RCW38}}$ 
includes planetary brightness temperature, angular integral, and responsivity 
uncertainties as well as the statistical error in the average flux between channels
of a common frequency.  }
\end{table*}

\begin{table*}
\begin{center}
\caption{Potential Sources of Systematic Error}
\label{tbl:systematics}
\begin{tabular}{|l|c|l|}
\hline
Source of error & Uncertainty & Mitigation \\
\hline\hline
Solid angle variation with chop & $\sim5\%$ & Measured with sources and accounted \\
 & & for in analysis. \\
\hline
Snow attenuation and chopper offsets & 4\% & LMT observing strategy.  Correlated \\
 & & with signal attenuation and cut. \\
\hline
Planetary flux consistency & 3\% & Cross-calibrated galactic sources with \\
 & & multiple planets. \\
\hline
Planetary voltage integral uncertainty & 2\% & No action taken. \\
\hline
Loading-induced detector responsivity & 2\% & Measured with calibrator source and \\
 changes & & corrected. \\
\hline
Measured passband & $\sim1\%$ & No action taken. \\
\hline
Baseplate drifts & $<1\%$ & Measured with thermometer and cut. \\
\hline
High-frequency leakage & $<1\%$ & No action taken. \\
\hline
Atmospheric opacity & $<0.5\%$ & Correlated measured opacity with \\
 & & 350$\mu$m tipper and corrected. \\
\hline
Detector time constant variation & $<0.1\%$ & No action taken. \\
\hline
RFI & - & Multiple layers of filtering and shielding. \\
\hline
Ground pickup & - & Primary guard ring, light baffle, \\
 & & telescope ground shield, LMT strategy. \\
\hline
\end{tabular}
\end{center}
\tablecomments{Table of possible sources of systematic errors.  We have included
an estimate of the potential effects upon the 150 GHz signal level or calibration
and listed the sources in order of severity. 
We also include any steps taken to mitigate these effects.  Table entries
denoted by a hyphen were not quantified. }
\end{table*}

\end{document}